\documentclass[12pt,prb,aps,showpacs,preprint]{revtex4}

\usepackage{graphics}
\usepackage{graphicx}
\usepackage{amsfonts}
\usepackage{textcomp}
\usepackage{amssymb}
\usepackage{amsmath}
\usepackage{color}
\usepackage{ulem}

\begin{document}

\title{Diversity of Dynamics and Morphologies of Invasive Solid Tumors}

\author{Yang Jiao}

\email{yjiao@princeton.edu}

\affiliation{Physical Science in Oncology Center,
Princeton University, Princeton New Jersey 08544, USA}

\author{Salvatore Torquato}

\email{torquato@electron.princeton.edu}

\affiliation{Department of Chemistry, Princeton University,
Princeton New Jersey 08544, USA}

\affiliation{Department of Physics, Princeton University,
Princeton New Jersey 08544, USA}

\affiliation{Physical Science in Oncology Center,
Princeton University, Princeton New Jersey 08544, USA}

\affiliation{Princeton Center for Theoretical Science, Princeton
University, Princeton New Jersey 08544, USA}

\affiliation{Program in Applied and Computational Mathematics,
Princeton University, Princeton New Jersey 08544, USA}

\date{\today}


\begin{abstract}
Complex tumor-host interactions can significantly affect the
growth dynamics and morphologies of progressing neoplasms. The
growth of a confined solid tumor induces mechanical pressure and
deformation of the surrounding microenvironment, which in turn
influences tumor growth. In this paper, we generalize a recently
developed cellular automaton model for invasive tumor growth in
heterogeneous microenvironments [Y. Jiao and S. Torquato, {\it
PLoS Comput. Biol.} {\bf 7}, e1002314 (2011)] by incorporating the
effects of pressure. Specifically, we explicitly model the
pressure exerted on the growing tumor due to the deformation of
the microenvironment and its effect on the local tumor-host
interface instability. Both {\it noninvasive-proliferative} growth
and {\it invasive} growth with individual cells that detach
themselves from the primary tumor and migrate into the surrounding
microenvironment are investigated. We find that while {\it
noninvasive} tumors growing in ``soft'' homogeneous
microenvironments develop almost isotropic shapes, both high
pressure and host heterogeneity can strongly enhance malignant
behavior, leading to finger-like protrusions of the tumor surface.
Moreover, we show that individual invasive cells of an {\it
invasive} tumor degrade the local extracellular matrix at the
tumor-host interface, which diminishes the fingering growth of the
primary tumor. The implications of our results for cancer
diagnosis, prognosis and therapy are discussed.
\end{abstract}

\maketitle

\section{Introduction}


Tumor malignancy arises from many complex interactions occurring
between the tumor and its host microenvironment. \cite{Co98,
hanahan00} There is increasing evidence that the host
microenvironment can significantly affect neoplasm progression.
\cite{deisboeck01, fidler03, chaplain98, gevertz06, jana08,
jana09, anderson05, anderson06, gatenby96b, gatenby06, gatenby06b,
bellomo00, scalerandi01, scalerandi02, kim10, stein07, macklin07,
jiaoplos1} The growth of a confined solid tumor also produces
mechanical pressure, leading to deformation of the surrounding
microenvironment, which generally affects the growth dynamics of
the tumor. \cite{helm97} Such pressure can result in clinical
complications, especially in a confined region of space such as
the brain \cite{hogan02} and deform or even collapse the
intra-tumoral blood and lymphatic vessels.\cite{pa2002} It has
been hypothesized that pressure may also influence tumor
physiology, growth rate and morphology. \cite{helm97} Therefore,
understanding effects of pressure on tumor growth is important for
both fundamental cancer research and clinical practice.
\cite{helm97}

Mechanical interactions between a tumor and its microenvironment
is a topic of great interest. In the work of Helmlinger et al,
\cite{helm97} it was shown that pressure can significantly reduce
tumor growth rate and even inhibit tumor growth \textit{in vivo}.
Bru and Casero \cite{bru06} studied the effect of external
pressure on the growth of tumor cell colonies and showed that
tumor morphology strongly depends on the pressure exerted by the
surrounding medium. Moreover, it has been observed that stiffer
microenvironments can promote malignant behavior.\cite{paszek05}
For example, tumors embedded in low-density soft agarose gels
remain roughly spherical in shape; however, they could exhibit a
finger-like morphology in a stiff gel with high density.
\cite{guiot07}


A variety of analytical and computational models have been
developed to incorporate the effect of pressure on tumor growth.
In particular, McElwain and Pettet \cite{mcelwain93} considered
that tumor cells mechanically behave as incompressible ``bags'' of
fluid enclosed by the plasma membrane. Chen et al. \cite{chen01}
modeled the growth of tumor spheroids in agarose gels, considering
the agarose gel to be a elastic material undergoing large
deformations and the tumor tissue was approximated by a fluid-like
material with additional drag and surface tension effects. Roose
et al. \cite{roose03} employed a linear poroelasticity model to
estimate the solid stress generated by the growth of the tumor
spheroid. Although the overall growth of the tumor spheroid can be
well described by these analytical approaches, they are not able
to provide detailed information on the tumor morphology. Gevertz
et al. \cite{jana08, jana09} employed a cellular automaton model
to investigate the effects the shape of an organ on growing tumors
through mechanical interactions. Using coupled nonlinear partial
differential equations, Macklin and Lowengrub \cite{macklin07}
modeled the response of the tissue surrounding the tumor to the
proliferation-induced mechanical pressure. Specifically, these
authors found that tumors growing in mechanically unresponsive
(i.e., rigid) microenvironments develop invasive fingering
morphologies and tumors growing in mechanically responsive (i.e.,
soft) microenvironment develop compact morphologies. However,
since a continuum method is used, it is not possible to keep track
of individual invasive cells that detach themselves from the
primary tumor and how such invasive cells affect the growth
dynamics and morphologies of the primary tumor.


Recently, we presented a single-cell based cellular automaton (CA)
model for invasive tumor growth in heterogeneous microenvironments
\cite{jiao11} in response to the challenge of developing an
``Ising'' model for cancer growth. \cite{To11} In this CA model,
individual invasive cells can detach themselves from the primary
tumor, locally degrade the extracellular matrix (ECM) and invade
into the surrounding host microenvironment. A rich spectrum of
emergent properties and coupled growth dynamics of the primary
tumor and invasive cells were predicted. However, the effects of
pressure exerted by the outer boundary of the growth permitting
region (e.g., cranium) on the tumor and the deformation of the ECM
were only implicitly considered.

In this paper, we generalize the aforementioned CA model to
explicitly take into account the deformation of the ECM
surrounding an invasive or noninvasive tumor, which in turn
imposes pressure on the neoplasm. Moreover, we also explicitly
consider the local geometry of the tumor-host interface (i.e., the
tumor surface), which can either enhance or reduce local growth
(i.e., interface instability) depending on the local curvature of
the interface. Both {\it noninvasive-proliferative} growth and
{\it invasive} growth with individual cells that detach themselves
from the primary tumor and migrate into the surrounding
microenvironment are investigated. We show here that by varying
the ECM rigidity (density), one can obtain a continuous spectrum
of tumor morphologies ranging from smooth isotropic shapes to
fingering patterns, which have been observed both \textit{in
vitro} and \textit{in vivo} \cite{macklin07}. The specific surface
\cite{torquato} is employed to quantify the degree of
``fingering'' for noninvasive proliferative growth. We find that
both the high pressure built up due to tumor growth and the
microenvironment heterogeneity can significantly promote
malignancy of the noninvasive proliferative tumor. Moreover, we
show that individual invasive cells that leave an invasive primary
tumor degrade the local ECM at the tumor-host interface, which
diminishes the fingering growth of the primary tumor. Our results
concerning the diversity of tumor morphologies enable one to infer
what are the possible mechanisms behind the resulting shapes. Such
information is expected to be of great value for cancer diagnosis,
prognosis and therapy.

\section{Computational Methods}

Following Refs.~[\onlinecite{kansal00a, kansal00b, kansal02,
jana08, jana09, jiao11}], we use the Voronoi tessellation
associated with random-sequential-addition (RSA) sphere packings
\cite{torquato} to model the underlying cellular structure. In
particular, nonoverlapping $d$-dimensional spheres ($d=2$ and 3)
are randomly and sequentially placed in a prescribed region in
$d$-dimensional Euclidean space until there is no void space left
for additional spheres. Then, space is divided into polyhedra,
each associated with a sphere center, such that any points within
a polyhedron is closer to its associated sphere center than to any
other sphere centers. The resulting Voronoi polyhedra are referred
to as automaton cells, which can represent either real biological
cells or regions of tumor stroma.

Here we explicitly takes into account the interactions between a
single cell and the surrounding microenvironment. Thus, each
automaton cell represents either a \textit{single} tumor cell
(approximately $15 - 20~\mu$m in size) or a region of tumor stroma
of similar size. In the current model, we mainly focus on the
effects of the ECM macromolecule density, ECM degradation by the malignant
cells, and the pressure due to the ECM deformation on tumor growth.
Henceforth, we will refer to the host microenvironment (or tumor
stroma) as the ``ECM'' for simplicity. Each ECM associated automaton
cell is assigned a particular density $\rho_{\mbox{\tiny{ECM}}}$,
representing the density of the ECM molecules within the automaton
cell. A tumor cell can occupy an ECM associated automaton cell
only if the density of this automaton cell
$\rho_{\mbox{\tiny{ECM}}} = 0$, which means that either the ECM is
degraded or it is deformed (pushed away) by the proliferating
tumor cells.

\subsection{Modeling the Pressure Exerted on the Growing Tumor}

The extracellular matrix is a complex mixture of macromolecules and
interstitial fluids that provides mechanical supports for the
tissue and plays an important role for cell adhesion and motility.
\cite{degrad1} In general, the ECM can be highly heterogeneous,
with large spatial variations of the ECM macromolecule densities
$\rho_{\mbox{\tiny ECM}}$. Our simulated tumors are only allowed
to grow in a compact growth-permitting region in order to mimic
the physical confinement of the host microenvironment, such as the
boundary of an organ or cranium in the case of the brain.
Therefore, a growing tumor deforms the ECM, which in turn
imposes a pressure on the tumor. \cite{helm97, mech1, mech2}

In Ref.~\onlinecite{jiao11}, we considered the effects of the local
ECM density on the proliferating cells. In this work, the ECM with
larger density was considered to be more rigid and more difficult
to degrade/deform. Therefore, the probability of division
$p_{div}$, which is related to the cell doubling time $\tau_0$ by
$\tau_0 = \ln 2 /\ln (1+p_{div})$, was taken to be a monotonically
decreasing function of $\rho_{\mbox{\tiny{ECM}}}$, e.g.,
\begin{equation}
\label{eq_pECM}
p_{div} \sim (1-\rho_{\mbox{\tiny{ECM}}}).
\end{equation}
The effect of pressure was only considered implicitly, e.g.,
$p_{div} \sim (1-r/L_{max})$, where $r$ is the distance of the
dividing cell from the tumor centroid, $L_{max}$ is the distance
between the closest growth-permitting boundary cell in the
direction of tumor growth and the tumor centroid.

Here we explicitly consider the pressure exerted on the growing
tumor by the ECM due to deformation. In particular, we consider
that the ECM is a linear elastic medium with bulk modulus
$\kappa_{\mbox{\tiny{ECM}}}$. The pressure $P$ due to volume
deformation $\Delta V$ is given by
\begin{equation}
\label{eq_pressure}
P = \kappa_{\mbox{\tiny{ECM}}} \frac{\Delta V}{V},
\end{equation}
where $V$ is the initial volume of the ECM. The ECM density
$\rho_{\mbox{\tiny{ECM}}} = M/V$, where $M$ is the ECM mass.
Therefore, the ECM density increase due to small volume shrinkage,
i.e.,
\begin{equation}
\label{eq_rho}
\rho_{\mbox{\tiny{ECM}}} = M/(V-\Delta V) \approx (M/V)(1+\Delta V) = \rho^0_{\mbox{\tiny{ECM}}} (1+\Delta V),
\end{equation}
which gives
\begin{equation}
\label{eq_dV}
\frac{\Delta V}{V} = \frac{(\rho_{\mbox{\tiny{ECM}}} - \rho^0_{\mbox{\tiny{ECM}}})}{\rho^0_{\mbox{\tiny{ECM}}}}.
\end{equation}
Substituting Eq.~(\ref{eq_dV}) into Eq.~(\ref{eq_pressure}), we
have
\begin{equation}
P = \kappa_{{\mbox{\tiny{ECM}}}} \frac{\Delta V}{V} = \kappa_{\mbox{\tiny{ECM}}}
\frac{(\rho_{\mbox{\tiny{ECM}}} - \rho^0_{\mbox{\tiny{ECM}}})}{\rho^0_{\mbox{\tiny{ECM}}}}.
\end{equation}
In other words, the pressure exerted by the ECM due to deformation
is proportional to its density, i.e., $P \sim
(\rho_{\mbox{\tiny{ECM}}}-\rho^0_{\mbox{\tiny{ECM}}})$. Without
loss of generality, we consider that cell division probability is
simply a monotonically decreasing function of pressure, and thus,
also a monotonically decreasing function of the ECM density, i.e.,
\begin{equation}
\label{eq_pP} p_{div} = (1-\omega P) \sim \left [{1- \omega
\frac{\kappa_{\mbox{\tiny{ECM}}}}{\rho^0_{\mbox{\tiny{ECM}}}}(\rho_{\mbox{\tiny{ECM}}}-\rho^0_{\mbox{\tiny{ECM}}})}\right]
= [1 - \frac{\omega^*}{\rho^0_{\mbox{\tiny{ECM}}}}
(\rho_{\mbox{\tiny{ECM}}}-\rho^0_{\mbox{\tiny{ECM}}})],
\end{equation}
where
\begin{equation}
\label{eq_omega} \omega^* = \omega \kappa_{\mbox{\tiny{ECM}}}
\end{equation}
is a constant of proportionality.

Suppose that the proliferative cells possess the ECM degradation
ability $\chi_0$, which is the fraction of the ECM macromolecules
degraded by malignant cells per day per unit volume. After each
day, the total mass of the ECM that has been degraded is
\begin{equation}
\Delta M = \chi_0 \sum_i^n \rho_{{\mbox{\tiny{ECM}}}}(i) v(i),
\end{equation}
where $n$ is the total number of the ECM associated automaton cells
taken by new tumor cells, $\rho_{\mbox{\tiny{ECM}}}(i)$ and $v(i)$
are respectively the macromolecule density and volume associated
with the $i$th automaton cell. The average ECM density is then given
by
\begin{equation}
\rho_{\mbox{\tiny{ECM}}} = \frac{M-\chi_0 \sum_i^n
\rho_{{\mbox{\tiny{ECM}}}}(i) v(i)}{V-\sum_i^n v(i)}.
\end{equation}
We define the ratio of
$\rho_{\mbox{\tiny{ECM}}}/\rho^0_{\mbox{\tiny{ECM}}}$ to be $\xi$,
i.e.,
\begin{equation}
\label{eq_xi} \xi =
\frac{\rho_{\mbox{\tiny{ECM}}}}{\rho^0_{\mbox{\tiny{ECM}}}} =
\frac{M-\chi_0 \sum_i^n
\rho_{{\mbox{\tiny{ECM}}}}(i)v(i)}{V-\sum_i^n v(i)} \frac{V}{M}.
\end{equation}
The macromolecule densities of the remaining ECM automaton cells
are then updated as
\begin{equation}
\label{eq_updaterho} \rho_{\mbox{\tiny{ECM}}}(j) = \xi
\rho^0_{\mbox{\tiny{ECM}}}(j),
\end{equation}
i.e., the increase of the ECM density after deformation is
proportional to its original density. Substitute
Eq.~(\ref{eq_updaterho}) into Eq.~(\ref{eq_pP}), we have
\begin{equation}
p_{div} \sim \left [{1- \omega^*(\xi-1)}\right].
\end{equation}
We note that in the above analysis, we have neglected the deformation
of the tumor cells, which possess a much larger bulk modulus
$\kappa_{\mbox{\tiny{cell}}}$ than that of the ECM, i.e.,
$\kappa_{\mbox{\tiny{cell}}}/\kappa_{\mbox{\tiny{ECM}}} \sim 100$
(see Ref.~\onlinecite{roose03}).

\subsection{Modeling Local Tumor-Host Interface Instablity}

Real tumors never possess a perfect spherical shape. Tumors
growing even in a homogeneous soft ECM will develop a ``bumpy'' tumor
surface, which can be very well captured by our underlying
cellular structure model (i.e., the Voronoi tessellation).
\cite{jiao11} When growing in a rigid microenvironment, a locally
smooth tumor surface which results in a
huge pressure gradient at the surface is highly undesirable. On the other hand, locally
small protrusions on the tumor surface can gain some growth
advantage by further invading into the surrounding ECM to release
local pressure. \cite{guiot07}

To model the aforementioned effects, we consider the local
geometry of the protrusion tip. In particular, the width of the
tip is taken to be the length $w$ of the automaton cell at the tip. The
length of the tip is given by
\begin{equation}
\ell = |{\bf x}_c - \overline{\bf x}|,
\end{equation}
where ${\bf x}_c$ is the position of the center of the automaton
cell at the tip and $\overline{\bf x} = \sum_i^m {\bf x}_i$ is the
average center position of tumor cells neighboring the cell at the
tip. The growth advantage of the cell at the tip is then
proportional to $\ell/w$, i.e.,
\begin{equation}
\label{eq_plw}
p_{div}\sim (1 + \ell/w).
\end{equation}
We note that $\ell$ is effective the radius of curvature, which
can be either positive or negative. A negative value of $\ell$
reduces $p_{div}$. For positive $\ell$, the ratio $\ell/w$ is
defined as the stress concentration factor associated with a crack
tip in solid mechanics.


Other biophysical mechanisms associated with noninvasive and
invasive malignant cells are the same as those described in
Ref.~\onlinecite{jiao11}. For example, the non-invasive cells
remain in the primary tumor and can be proliferative, quiescent or
necrotic, depending on their nutritional supply, which is
determined by tumor-size dependent characteristic diffusion
distances. Proliferative cells can produce ``mutant'' daughter
cells that possess strong ECM degradation ability $\chi_1$ and can
leave the primary tumor and invade into the surrounding
microenvironment by locally degrading the ECM macromolecules. Readers
are referred to Ref.~\onlinecite{jiao11} for details of such
mechanisms.

\subsection{Cellular Automaton Rules}

We now specify the CA rules for our generalized model, which
closely follow those given in Ref.~\onlinecite{jiao11}, except for
the additional rules explicitly incorporating the pressure imposed
by the ECM and the local host-tumor interface instability described
here. After generating the automaton cells using Voronoi
tessellation, an ECM density $\rho_{\mbox{\tiny{ECM}}} \in (0,
~1)$ is assigned to each automaton cell within the
growth-permitting region, which represents the heterogeneous host
microenvironment. Then a tumor is introduced by designating any
one or more of the automaton cells as proliferative cancer cells.
Time is then discretized into units that represent one real day.
At each time step:

\begin{itemize}

\item Each automaton cell is checked for type: invasive,
proliferative, quiescent, necrotic or ECM associated. Invasive
cells degrade and migrate into the ECM surrounding the tumor.
Proliferative cells are actively dividing cancer cells, quiescent
cancer cells are those that are alive, but do not have enough
oxygen and nutrients to support cellular division and necrotic
cells are dead cancer cells.

\item All tumorous necrotic cells are inert (i.e., they do not
change type).

\item Quiescent cells more than a certain distance $\delta_n$ from
the tumor's edge are turned necrotic. The tumor's edge, which is
assumed to be the source of oxygen and nutrients, consists of all
ECM associated automaton cells that border the neoplasm. The
critical distance $\delta_n$ for quiescent cells to turn necrotic
is computed as follows:
\begin{equation}
\label{eq_delta} \delta_n = aL_t^{(d-1)/d},
\end{equation}
where $a$ is a prescribed parameter (see Table \ref{tab_Param}), $d$
is the Euclidean spatial dimension and $L_t$ is the distance between the
geometric centroid ${\bf x}_c$ of the tumor and the tumor edge
cell that is closest to the quiescent cell under consideration.
The position of the tumor centroid ${\bf x}_c$ is given by
\begin{equation}
\label{eq_xc} {\bf x}_c = \frac{{\bf x}_1 + {\bf x}_2 + \cdots +
{\bf x}_N}{N},
\end{equation}
where $N$ is the total number of noninvasive cells contained in
the tumor, which is updated when a new noninvasive daughter cell
is added to the tumor.

\item Each proliferative cell will attempt to divide with
probability $p_{div}$ into the surrounding ECM (i.e., the
automaton cells associated with the ECM) by degrading and pushing
away the ECM in that automaton cell. As discussed in the previous
section, we consider that $p_{div}$ for a specific proliferative
cell depends on the local ECM density [Eq.~(\ref{eq_pECM})], the
pressure imposed by the ECM [Eq.~(\ref{eq_pP})] and the local geometry of the
tumor-host interface [Eq.~(\ref{eq_plw})], i.e.,
\begin{equation}
\label{eq_pdiv} p_{div} = \left\{
\begin{array}{ll}
{p_0}[1-\rho_{\mbox{\tiny{ECM}}}-\omega^*\xi & \mbox{if any ECM associated automaton cell within} \\
\quad\quad\quad +\omega^* + \xi\frac{\ell}{w}] & \mbox{the predefined growth distance is in the growth-}\\
 & \mbox{permitting microenvironment} \\  \\
 & \mbox{if no ECM associated automaton cell within} \\
0 & \mbox{the predefined growth distance is in the growth-}\\
& \mbox{permitting microenvironment, or the value}\\
& \mbox{of the above expression is negative}
\end{array}
\right.
\end{equation}
where $p_0$ is the base probability of division (see Table
\ref{tab_Param}), $r$ is the distance of the dividing cell from
the tumor centroid, $\rho_{\mbox{\tiny{ECM}}}$ is the ECM density
of the automaton cell to be taken by the new tumor cell,
$\omega^*$ and $\xi$ are respectively given by
Eq.~(\ref{eq_omega}) and Eq.~(\ref{eq_xi}). When a ECM associated
automaton cell is taken by a tumor cell, its density is set to be
zero. The predefined growth distance ($\delta_p$) is described in
the following bullet point.

\item If a proliferative cell divides, it can produce a mutant
daughter cell possessing an invasive phenotype with a prescribed
probability $\gamma$ (i.e., the mutation rate). The invasive
daughter cell gains ECM degradation ability $\chi_1$ and motility
$\mu$, which enables it to leave the primary tumor and invade into the
surrounding ECM. The rules for updating invasive cells are given
in the following bullet point. If the daughter cell is
noninvasive, it is designated as a new proliferative cell.

\item A proliferative cell turns quiescent if there is no space
available for the placement of a daughter cell within a distance
$\delta_p$ from the proliferative cell, which is given by
\begin{equation*}
\label{eq_deltap} \delta_p = bL_t^{(d-1)/d},
\end{equation*}
where $b$ is a nutritional parameter (see Table \ref{tab_Param}), $d$
is the spatial dimension and $L_t$ is the distance between the
geometric tumor centroid ${\bf x}_c$ and the tumor edge cell that
is closest to the proliferative cell under consideration.

\item An invasive cell degrades the surrounding ECM (i.e., those
in the neighboring automaton cells of the invasive cell) and can
move from one automaton cell to another if the associated ECM is
completely degraded locally. For an invasive cell with motility
$\mu$ and ECM degradation ability $\chi_1$, it will make $m$
attempts to degrade the ECM in the neighboring automaton cells and
jump to these automaton cells, where $m$ is an arbitrary integer
in $[0,~\mu]$. For each attempt, the surrounding ECM density
$\rho_{\mbox{\tiny{ECM}}}$ is decreased by $\delta\rho$, where
$\delta\rho$ is an arbitrary number in $[0,~\chi_1]$. Using random
numbers for the ECM degradation ability and cellular motility is to
take into account tumor genome heterogeneity, which is manifested
as heterogeneous phenotypes (such as different $m$ and
$\delta\rho$). When the ECM in multiple neighboring automaton
cells of the invasive cell are completely degraded (i.e.,
$\rho_{\mbox{\tiny{ECM}}} = 0$), the invasive cell moves in a
direction that maximizes the nutrients and oxygen supply. Here we
assume that the migrating invasive cells \textit{do not divide}. The
degraded ECM shows the invasive path of the tumor.

\item The density $\rho_{\mbox{\tiny{ECM}}}$ of the remaining ECM
automaton cells is updated according to Eq.~(\ref{eq_updaterho}).

\end{itemize}

\begin{table}
\caption{Parameters and terms in the CA model. Summarized here are
definitions of the parameters for tumor growth and invasion, and
all other (time-dependent) quantities used in the simulations.
The number(s) listed in parentheses indicates the
value or range of values assigned to the corresponding parameters in the
simulations. The values of the parameters are chosen such that the
CA model can reproduce reported growth dynamics of GBM from the
medical literature \cite{deisboeck01, kansal00a, jana08}.}
\begin{center}
\begin{tabular}{ll} \\ \hline\hline
\multicolumn{2}{c}{\textbf{Time dependent terms}} \\
\hline
$L_t$ & Local tumor radius (varies with cell positions) \\
\hline
$L_{max}$ & Local maximum tumor extent (varies with cell positions) \\
\hline
$\delta_p$ & Characteristic proliferative rim thickness \\
\hline
$\delta_n$ & Characteristic living-cell rim thickness (determines necrotic fraction) \\
\hline
$p_{div}$ &
Probability of division (varies with cell positions) \\
\hline
$\rho_{\mbox{\tiny{ECM}}}$ & ECM density (depends on ECM deformation and varies with positions, $>0$) \\
\hline
$\xi$ & Ratio of current ECM density over initial density
$\rho_{\mbox{\tiny ECM}}/\rho^0_{\mbox{\tiny ECM}}$ \\
\hline
$\omega^*$ &Parameter measuring $p_{div}$ reduced by pressure,
$=2\rho^0_{\mbox{\tiny ECM}}$\\ 
\hline\hline \multicolumn{2}{c}{\textbf{Growth parameters}} \\
\hline
$p_0$ & Base probability of division, linked to cell-doubling time (0.192) \\
\hline
$a$ & Base necrotic thickness, controlled by nutritional needs ($0.58$ mm$^{1/2}$) \\
\hline
$b$ & Base proliferative thickness, controlled by nutritional needs ($0.30$ mm$^{1/2}$) \\
\hline
$\ell$ & Length of local protrusion tip \\
\hline
$w$ & Width of local protrusion tip \\
\hline
$\chi_0$ & ECM degradation ability of proliferative cells ($0.0-0.25$)\\
\hline\hline
\multicolumn{2}{c}{\textbf{Invasiveness parameters}}
\\ \hline
$\gamma$ & Mutation rate (determines the number of invasive cells, 0.05) \\
\hline
$\chi_1$ & ECM degradation ability of invasive cells ($0.4-1.0$) \\
\hline
$\mu$ & Cell motility (the number of ``jumps'' from one automaton cell to another, $0-4$) \\ 
\hline\hline
\end{tabular}
\end{center}
\label{tab_Param}
\end{table}

The important parameters mentioned in the bullet points above are
summarized in Table \ref{tab_Param}. We note that although only
spherical growth-permitting regions are considered here, this constraint can
be easily relaxed. As a demonstration of the capability and
versatility of the generalized CA model, we will employ it to
investigate the growth dynamics and morphologies of both
noninvasive and invasive tumors in two dimensions. However, the
model is easily extended to three dimensions and the algorithmic
details of the model are presented for any spatial dimension.

\section{Results}

Homogeneous and random distributions of the ECM density
\cite{jiao11} are used to study the effects of microenvironment
heterogeneity on the growing tumor. The random distribution
of the ECM density, which henceforth is referred to as the ``random ECM''
for simplicity, is generated by assigning a random ECM density
value between 0 and 1 to each ECM associated automaton cell. The
boundary of the growth-permitting region is considered to be
vascularized, i.e., a growing tumor can receive oxygen and
nutrients from the growth-permitting region. In particular, we
consider a constant radially symmetric nutrient/oxygen gradient in
the growth-permitting region with the highest nutrient/oxygen
concentration at the vascular boundary. We note that although
generally the nutrient/oxygen concentration field \textit{in vivo} is more
complicated, previous numerical studies that considered the exact
evolution of nutrient/oxygen concentrations have shown a decay of
the concentrations toward the tumor center. \cite{anderson05,
anderson06} Since the directions of cell motions are determined by
the nutrient/oxygen gradient only, our constant-gradient
approximation is a very reasonable one.

In the beginning, a proliferative tumor cell is introduced at the
center of the growth-permitting region and tumor growth is
initiated. The parameters employed are either given in Table
\ref{tab_Param} or specified for each case separately. The
specific surface $s$ for the noninvasive proliferative tumor, defined as the ratio
of the total length of the perimeter of the primary tumor over its
total area, \cite{torquato, jiao11} is employed to quantify the
degree of ``fingering'' of the growing tumor. For a perfectly
circular shape with radius $R$, the associated $s$ is given by $2/R$,
which is the minimal value among all shapes with the same area. The
specific surface of a tumor in excess of that of a circle
provides a measurement of the roughness of the tumor surface and thus,
the degree of ``fingering''. Therefore, the specific surface $s$
scaled by $2/R_T$ associated with a circle is
used for an arbitrary-shaped tumor with effective radius $R_T$
(i.e., the average distance from tumor edge to tumor center). In the visualizations
of the tumor that follow, we will use the following convention: The ECM
degraded by the tumor cells is blue. In the primary tumor,
necrotic cells are black, quiescent cells are yellow and
proliferative cells are red. The invasive tumor cells are green.

\subsection{Noninvasive Proliferative Growth}

We first investigate the effects of pressure on the growth
of noninvasive proliferative tumors by setting the mutation rate to zero,
i.e., $\gamma = 0$. Figure 1 shows the snapshots of noninvasive tumors
growing in the homogeneous ECM with initial density
$\rho^0_{\mbox{\tiny ECM}} = 0.25, 0.45$ and 0.65. The associated
specific surface as a function of time is also shown. The ECM
degradation ability value $\chi_0 = 0.2$ for the proliferative cells is used.

It can be clearly seen that as the ECM density (i.e., pressure
level) varies, a variety of growth dynamics and tumor morphologies
emerge. In particular, the tumor growing in the ECM with low
density (pressure) ($\rho_{\mbox{\tiny ECM}} = 0.25$)
develops an almost isotropic shape with small specific surface $s/(2/R_T)$
(upper panel of Fig. 1). As the ECM density (pressure) increases
($\rho_{\mbox{\tiny ECM}} = 0.45$), the tumor begins to develop
bumpy surface, but a well-defined ``spherical'' core can still be
identified (middle panel of Fig. 1). For the high ECM density
(pressure) $\rho_{\mbox{\tiny ECM}} = 0.65$, finger-like
protrusions emerge at early growing stages and no ``spherical''
core of significant size is found (lower panel of Fig. 1).
The associated specific surface $s/(2/R_T)$ monotonically increases as the
tumor grows. Since the elongated finger-like structures of
proliferative cells can grow into the nutrient-rich
microenvironment, necrotic regions within these ``fingers'' are
very rare. We note that the overall size (i.e., distance between
tumor center and the farthest cell on tumor edge) of the tumors
with distinct morphologies decrease as the ECM density and the
associated pressure increase. Nonetheless, the emergence of
finger-like structures significantly release the local pressure built
up at the smooth surface of the growing tumor and thus, the
growth of such invasion fingers is favored.

Figure 2 shows the snapshots of noninvasive tumors (e.g., $\gamma = 0$)
growing in the random ECM with average initial density
$\overline\rho^0_{\mbox{\tiny ECM}} = 0.25, 0.45$ and 0.65, as
well as the associated specific surface $s/(2/R_T)$ as a function of time. The ECM
degradation ability value $\chi_0 = 0.2$ for the proliferative cells is
used. Similar effects of the ECM density (i.e., pressure level) on
the growth dynamics and tumor morphologies are observed, i.e.,
increasing the ECM density (pressure) leads to a continuous variation
of tumor morphology ranging from smooth isotropic shapes to
significantly fingered patterns. Since the heterogeneity of the ECM
results in stronger local tumor surface roughness, the fingering
effect is also stronger comparing with the corresponding
homogeneous case with the same ECM density. In particular, the
tumor growing in the ECM with $\overline\rho^0_{\mbox{\tiny ECM}} =
0.65$ develops sub-fingers on the primary fingers, which is mainly
caused by the local ECM density fluctuations (lower panel of Fig.
2). Such fine morphological features can hardly be resolved by
continuum simulation method. The stronger fingering of tumors growing in the
random ECM also leads to larger overall extents of the tumors.

\subsection{Invasive Growth with Individual Cells Migrating into Surround ECM}

Figure 3 shows the snapshots of invasive tumors growing in a
homogeneous ECM with initial density $\rho^0_{\mbox{\tiny ECM}} =
0.25, 0.45$ and 0.65 on day 120.
Specifically, individual invasive cells can detach themselves from the
primary tumor and migrate into the surrounding ECM.
The following values of
invasiveness parameters are used: mutation rate $\gamma = 0.05$,
ECM degradation ability value for the proliferative cells $\chi_0 = 0.25$,
ECM degradation ability value for the invasive cells $\chi_1 = 0.75$,
motility of the invasive cells $\mu = 4$, which corresponds to a high
degree of malignancy.

It is clear from Fig. 3 that the tumor growing in the low-density
ECM (e.g., under low pressure) ($\rho_{\mbox{\tiny ECM}} = 0.25$) develops
relatively short invasive branches (Fig. 3a), while the tumor
growing in the high-density ECM (e.g., under high pressure) ($\rho_{\mbox{\tiny
ECM}} = 0.65$) possesses very long invasive branches (Fig. 3c).
Also, we note that a large number of invasive cells are generated
at the finger tips of the primary tumor, which in turns promotes
the growth of the fingers. Since the invasive cells degrade the
ECM close to the tumor surface, the local pressure exerted on the
tumor is reduced. Thus, the degree of ``fingering'' in the
invasive tumors is smaller than in the noninvasive ones.
The invasive cells also enhance the growth of the primary tumor.
Moreover, it can be seen that high ECM density (i.e., high
pressure exerted on the tumor) enhances both fingering of the
primary tumor and malignant behavior of invasive cells (Fig. 3c).
We note that such invasion-tumor couplings have been extensively
explored in Ref.~\onlinecite{jiao11}.

Figure 4 shows the snapshots of invasive tumors growing in the random ECM
with average initial density $\overline\rho^0_{\mbox{\tiny ECM}} =
0.25, 0.45$ and 0.65 on day 120. The same values of invasiveness
parameters as the homogeneous case are used. Again, the growth
dynamics and tumor morphologies are qualitatively the same as the
corresponding homogeneous case. However, the heterogeneity of the
ECM enhances local tumor instability and thus, promotes the
malignant behavior of the invasive tumor.

\section{Conclusions and Discussion}


We have generalized a recently developed cellular automaton
model to study the effects of pressure on the
dynamics and tumor morphologies for both noninvasive proliferative
growth and invasive growth with individual malignant cells detaching
themselves from the primary tumor. In particular, we have explicitly
taken into account the deformation of the extracellular matrix
surrounding the tumor, which in turn imposes pressure on the
neoplasm. Moreover, we also considered the local tumor-host
interface instability, which can give rise to the emergence of
finger-like protrusions of the tumor surface. We showed that by
varying the ECM rigidity (i.e., the pressure level in the ECM), one can
obtain a variety of tumor morphologies ranging from spherical
shapes to fingering patterns, which are quantitatively
characterized by the specific surface. We also found that both high
pressure and microenvironment heterogeneity can amplify the
malignancy of the neoplasm in both the noninvasive proliferative case
and the invasive case. Moreover,
we demonstrated that the growth dynamics of the primary tumor and invasive cells
are strongly coupled.

Figure 5 shows the morphology of ductal carcinoma \textit{in situ} (DCIS),
which resembles the morphology of noninvasive tumors growing in
the ECM with intermediate density (Fig. 1(b) and Fig. 2(b))
predicted by our CA model. Strong fingering growth is rarely
observed for DCIS, manly because the neoplasm is further
constrained by a tight basal membrane composed of epithelial
cells. Such complex microenvironment heterogeneities need to be
incorporated in the CA model to accurately predict DCIS
progression. On the other hand, significant fingering has been
observed \textit{in vitro}.\cite{macklin07} These experimental observations clearly
demonstrate the robustness and predictive capability of our CA
model.

We have shown that a high pressure in the ECM, which is due to
large ECM density and deformation, can lead to significant
fingering growth and enhance the malignant behavior of both the
primary tumor and invasive cells. As a tumor grows in a
confined microenvironment, it is inevitable that a high pressure
will be built up. Thus, a tumor with a low level of malignancy
initially can eventually develop highly malignant invasive
behavior. Specifically, when finger-like protrusions develop,
cells close to the finger tip have less contacting neighbors and
thus, less adhesion. This makes it very easy for the invasive
cells to leave the primary tumor and migrate deeply into the
surrounding microenvironment, which ultimately leads to cancer
metastasis.

Moreover, our results concerning the diversity of tumor
morphologies enable one to infer the possible mechanisms behind
the resulting shapes. For example, a noninvasive
proliferative tumor possesses a morphology with significant
finger-like protrusions on the tumor surface could be attributed to a
host microenvironment in which the tumor grew that was very
rigid and inhomogeneous. On the other hand, if such a tumor has a
smooth and almost isotropic shape, it could mean that its host environment was
very soft and homogeneous. For an invasive tumor with individual
invasive cells that detach themselves from the primary tumor and
migrate into the surrounding microenvironment, a rougher tumor
surface could imply that the individual invasive cells possessed a
strong ECM degradation ability, high motility and weak cell-cell
adhesion.


Although our CA model is readily applied to model {\it in vitro} tumor growth,
the heterogeneous microenvironments considered in the current model
are highly idealized and do not include heterogeneities such as blood
vessels and lymphatics, which could play an important role in clinical cancers.
Incorporating more realistic microenvironments as well as other
possible mechanisms (e.g., tumor and normal cell phenotypic plasticity
and immune response) would lead to an improved model that could provide
insights into {\it in vivo} tumor growth. Nonetheless, we expect that the conclusions drawn here still
qualitatively apply to {\it in vivo} situations.
Therefore, information on the tumor morphology, which can be
obtained from histological images, is expected to lead to more accurate
diagnosis and thus, more effective tumor treatment strategies. For
example, if a tumor with a rough surface is detected, drugs that can
release the high concomitant pressure in the host environment by modifying the
molecular compositions of the ECM macromolecules could be used to
reduce the malignancy of the tumor, leading to a noninvasive
smooth isotropic shape. This not only could improve the efficiency
of chemotherapy but also could make it easier to remove the tumor
by resection. Finally, we note that improving the deliverability of chemotherapy
will require application of existing theories to predict the transport properties
of the underlying heterogeneous media.\cite{transport}


\begin{acknowledgments}
The research described was supported by the National Cancer
Institute under Award NO. U54CA143803. The content is solely the
responsibility of the authors and does not necessarily represent
the official views of the National Cancer Institute or the
National Institutes of Health.
\end{acknowledgments}

\clearpage

\newpage

\begin{figure}
\centering \caption{Upper panel: Snapshots of a noninvasive tumor
growing in homogeneous ECM with density $\rho_{\mbox{\tiny ECM}} =
0.25$ on day 40 (a1), day 80 (a2) and day 120 (a3) after
initialization. The associated specific surface $s/(2/R_T)$ (a4) remains small
in value as the tumor grows, indicating a compact tumor
morphology. Middle panel: Snapshots of a noninvasive tumor growing
in homogeneous ECM with density $\rho_{\mbox{\tiny ECM}} = 0.45$
on day 40 (b1), day 80 (b2) and day 120 (b3) after initialization.
The associated specific surface $s/(2/R_T)$ (b4) shows a rapid growth after
day 100, indicating fingering growth of the tumor. Lower panel:
Snapshots of a noninvasive tumor growing in homogeneous ECM with
density $\rho_{\mbox{\tiny ECM}} = 0.65$ on day 40 (c1), day 80
(c2) and day 120 (c3) after initialization. The associated
specific surface $s/(2/R_T)$ (c4) increase monotonically as the tumor grows,
indicating significant fingering of the tumor. } \label{fig1}
\end{figure}

\begin{figure}
\centering \caption{Upper panel: Snapshots of a noninvasive tumor
growing in random ECM with average density
$\overline\rho_{\mbox{\tiny ECM}} = 0.25$ on day 40 (a1), day 80
(a2) and day 120 (a3) after initialization. The associated
specific surface $s/(2/R_T)$ (a4) remains small in value as the tumor grows,
indicating a compact tumor morphology. Middle panel: Snapshots of
a noninvasive tumor growing in random ECM with average density
$\overline\rho_{\mbox{\tiny ECM}} = 0.45$ on day 40 (b1), day 80
(b2) and day 120 (b3) after initialization. The associated
specific surface $s/(2/R_T)$ (b4) shows a rapid growth after day 100,
indicating fingering growth of the tumor. Lower panel: Snapshots
of a noninvasive tumor growing in random ECM with average density
$\rho_{\mbox{\tiny ECM}} = 0.65$ on day 40 (c1), day 80 (c2) and
day 120 (c3) after initialization. The associated specific surface $s/(2/R_T)$
(c4) increase monotonically as the tumor grows, indicating
significant fingering of the tumor. Note that sub-fingers are
developed on the primary fingers.} \label{fig2}
\end{figure}

\begin{figure}
\centering \caption{Snapshots of invasive tumors growing in a
homogeneous ECM with different densities on day 100. Individual
invasive cells detach themselves from the primary tumor, locally
degrade the ECM and migrate into the surrounding microenvironment.
(a) $\rho_{\mbox{\tiny ECM}} = 0.25$ (b) $\rho_{\mbox{\tiny ECM}} =
0.45$ (c) $\rho_{\mbox{\tiny ECM}} = 0.65$.
Note that the tumor growing in the low-density
ECM (pressure) ($\rho_{\mbox{\tiny ECM}} = 0.25$) develops
relative short invasive branches [see panel (a)], while the tumor
growing in the high-density ECM (pressure) ($\rho_{\mbox{\tiny
ECM}} = 0.65$) possesses very long invasive branches [see panel (c)].} \label{fig3}
\end{figure}

\begin{figure}
\centering \caption{Snapshots of invasive tumors growing in a
random ECM with different average densities on day 100.
Individual invasive cells detach themselves from the primary tumor, locally
degrade the ECM and migrate into the surrounding microenvironment.
(a) $\overline\rho_{\mbox{\tiny ECM}} = 0.25$ (b)
$\overline\rho_{\mbox{\tiny ECM}} = 0.45$ (c)
$\overline\rho_{\mbox{\tiny ECM}} = 0.65$.
Note that the tumor growing in the low-density
ECM (pressure) ($\rho_{\mbox{\tiny ECM}} = 0.25$) develops
relative short invasive branches [see panel (a)], while the tumor
growing in the high-density ECM (pressure) ($\rho_{\mbox{\tiny
ECM}} = 0.65$) possesses very long invasive branches [see panel (c)].
Also observe that invasive cells clump
at the finger tips of the primary tumor, which in turns promotes
the growth of the fingers [see panel (c)].} \label{fig4}
\end{figure}

\begin{figure}
\centering \caption{An image showing the morphology of ductal
carcinoma in situ with bumpy surface. Image courtesy of R.
Gatenby.} \label{fig5}
\end{figure}

\clearpage
\newpage

\setcounter{figure}{0}

\begin{figure}
$\begin{array}{c@{\hspace{0.5cm}}c@{\hspace{0.5cm}}c@{\hspace{0.5cm}}c}
\includegraphics[height=3.5cm, keepaspectratio]{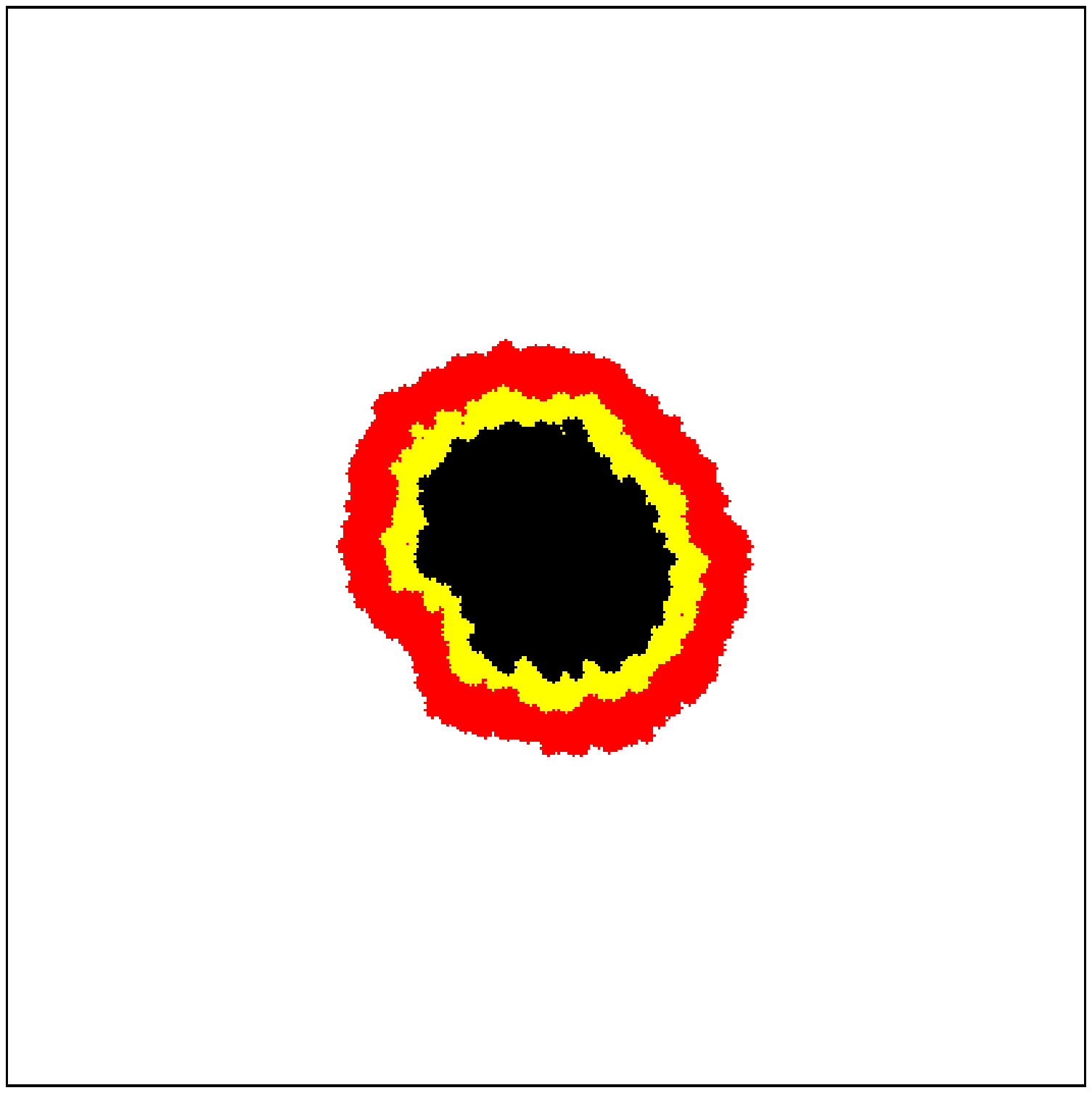} &
\includegraphics[height=3.5cm, keepaspectratio]{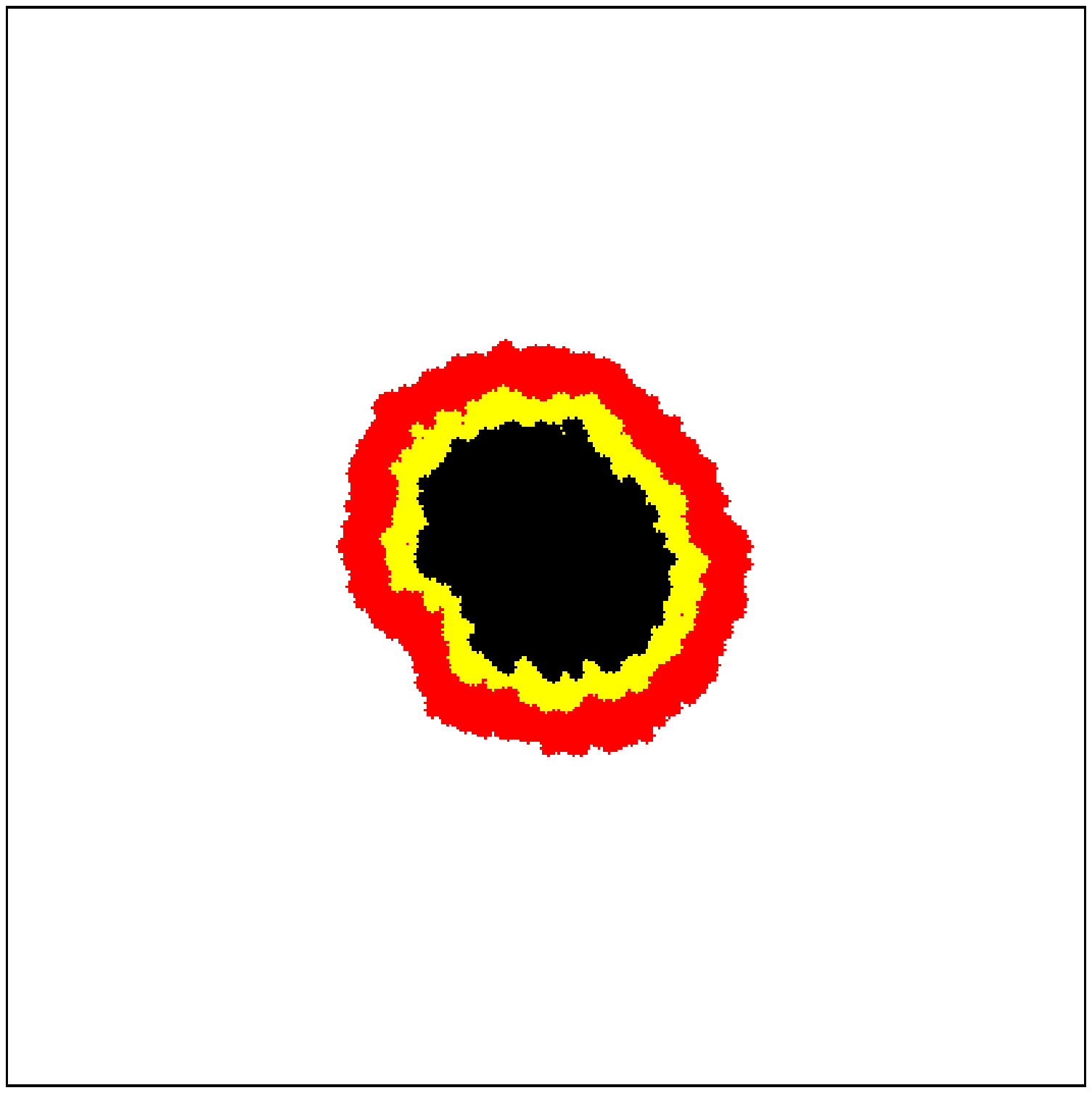} &
\includegraphics[height=3.5cm, keepaspectratio]{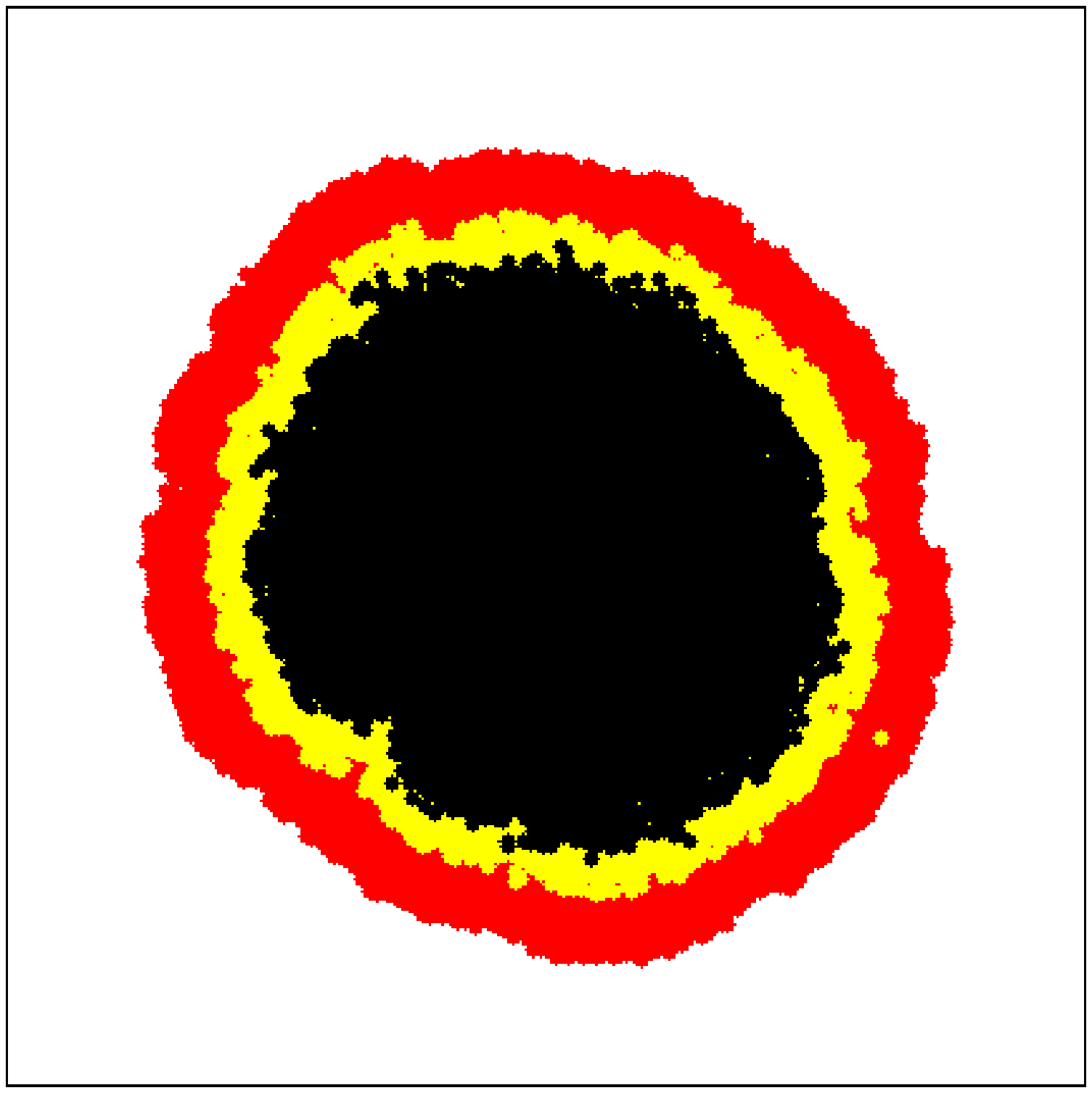} &
\includegraphics[height=3.5cm, keepaspectratio]{fig1a4.eps} \\
\mbox{(a1)} & \mbox{(a2)} & \mbox{(a3)} & \mbox{(a4)} \\\\
\includegraphics[height=3.5cm, keepaspectratio]{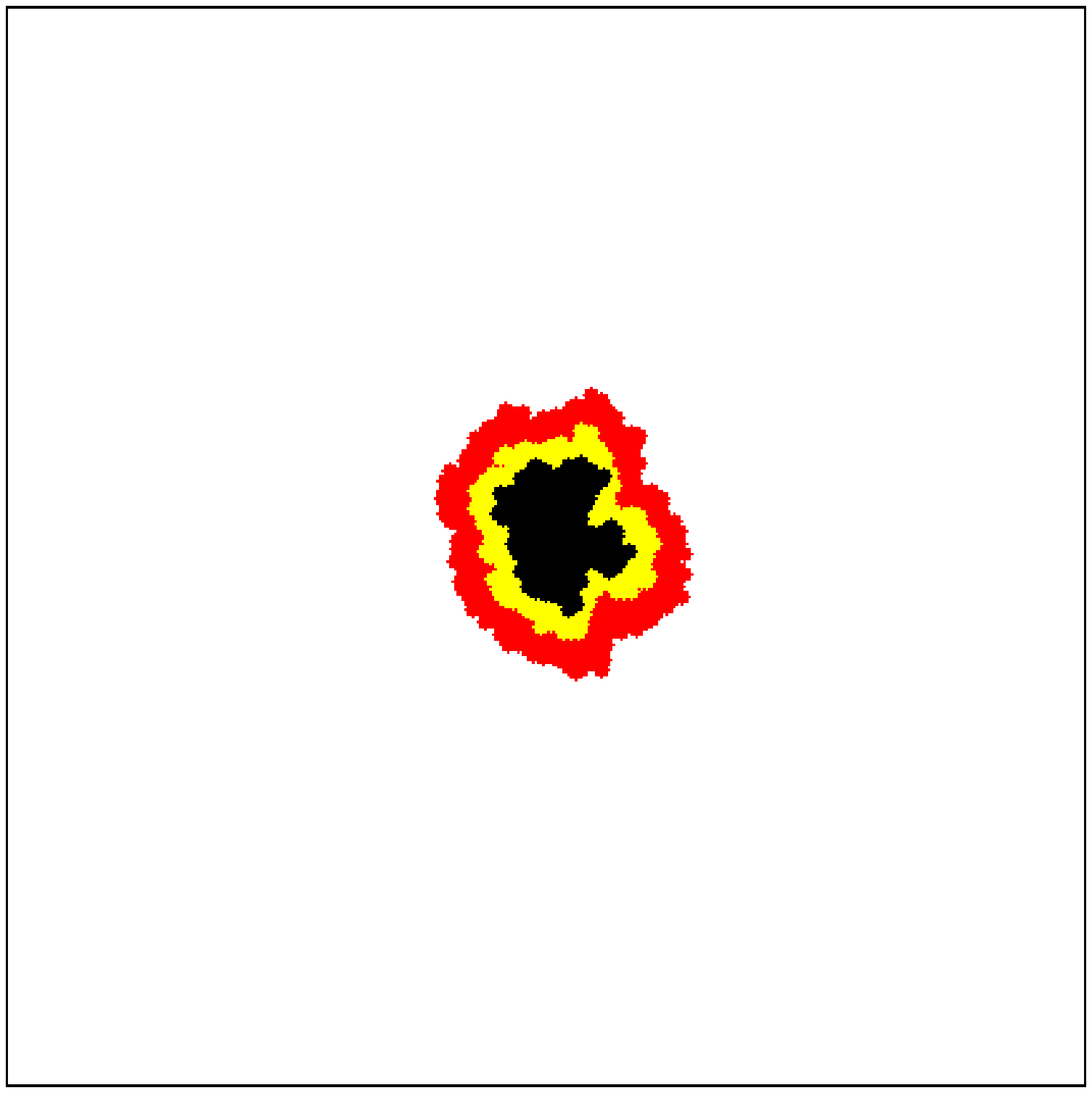} &
\includegraphics[height=3.5cm, keepaspectratio]{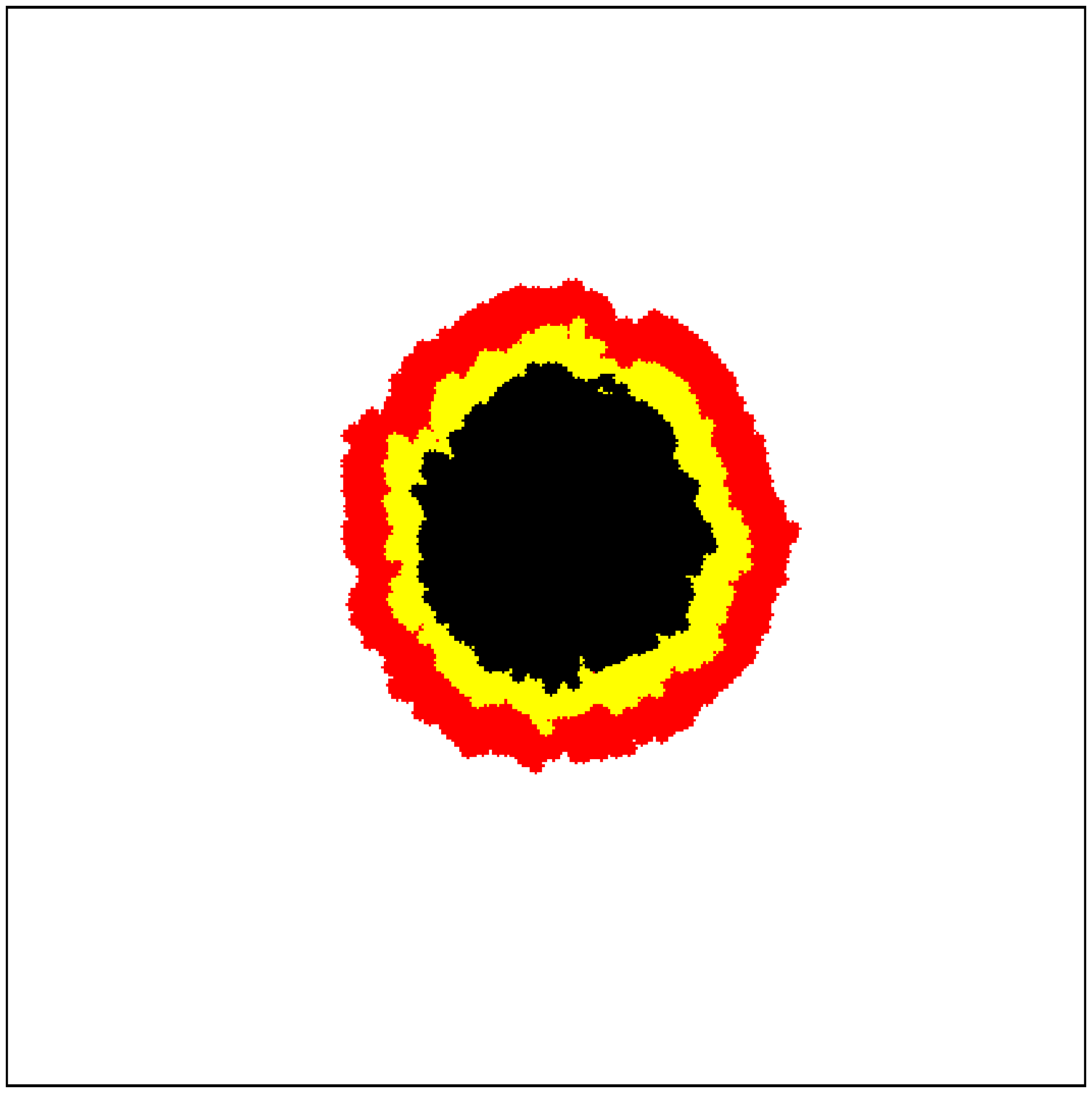} &
\includegraphics[height=3.5cm, keepaspectratio]{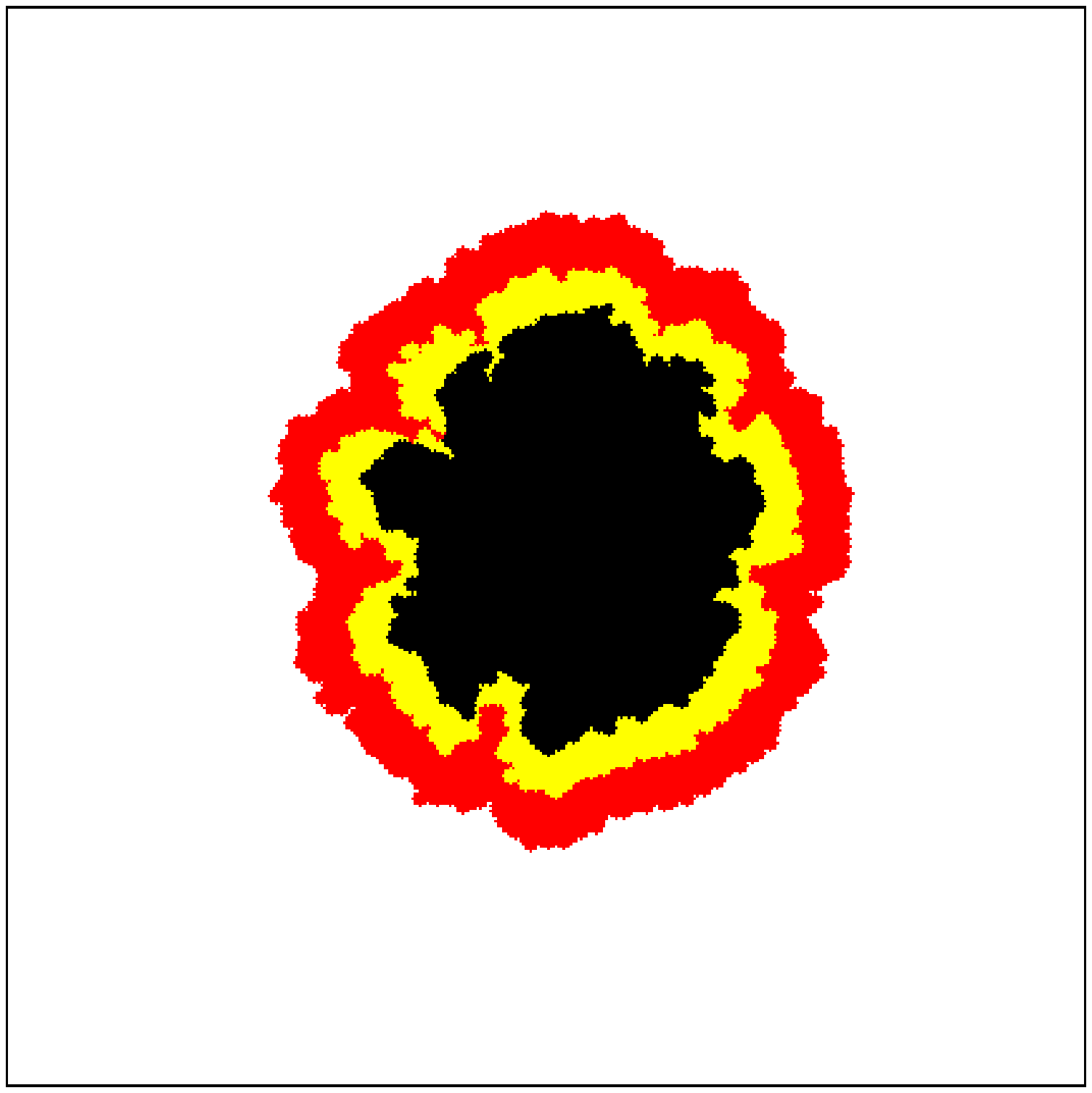} &
\includegraphics[height=3.5cm, keepaspectratio]{fig1b4.eps} \\
\mbox{(b1)} & \mbox{(b2)} & \mbox{(b3)} & \mbox{(b4)} \\\\
\includegraphics[height=3.5cm, keepaspectratio]{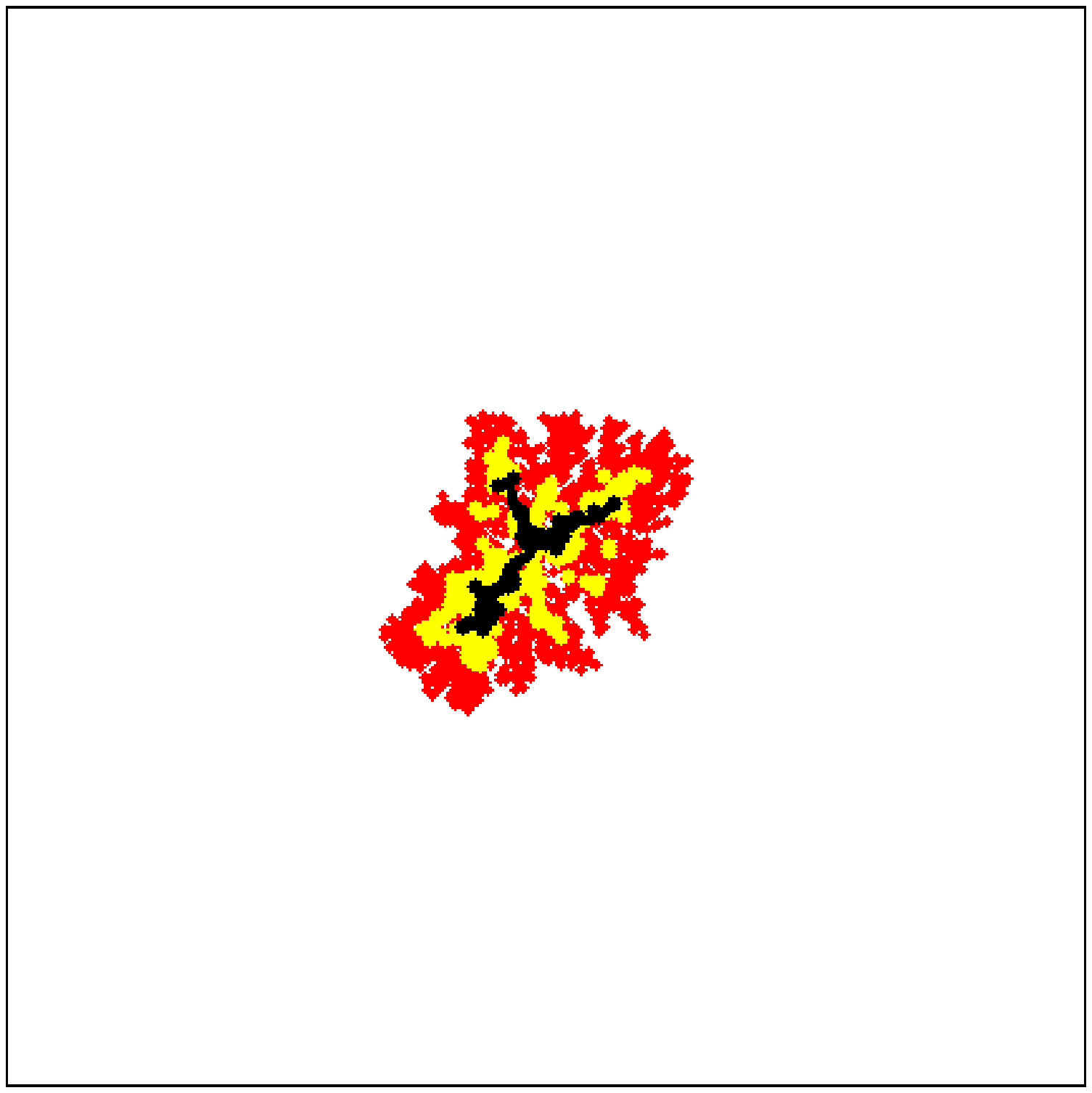} &
\includegraphics[height=3.5cm, keepaspectratio]{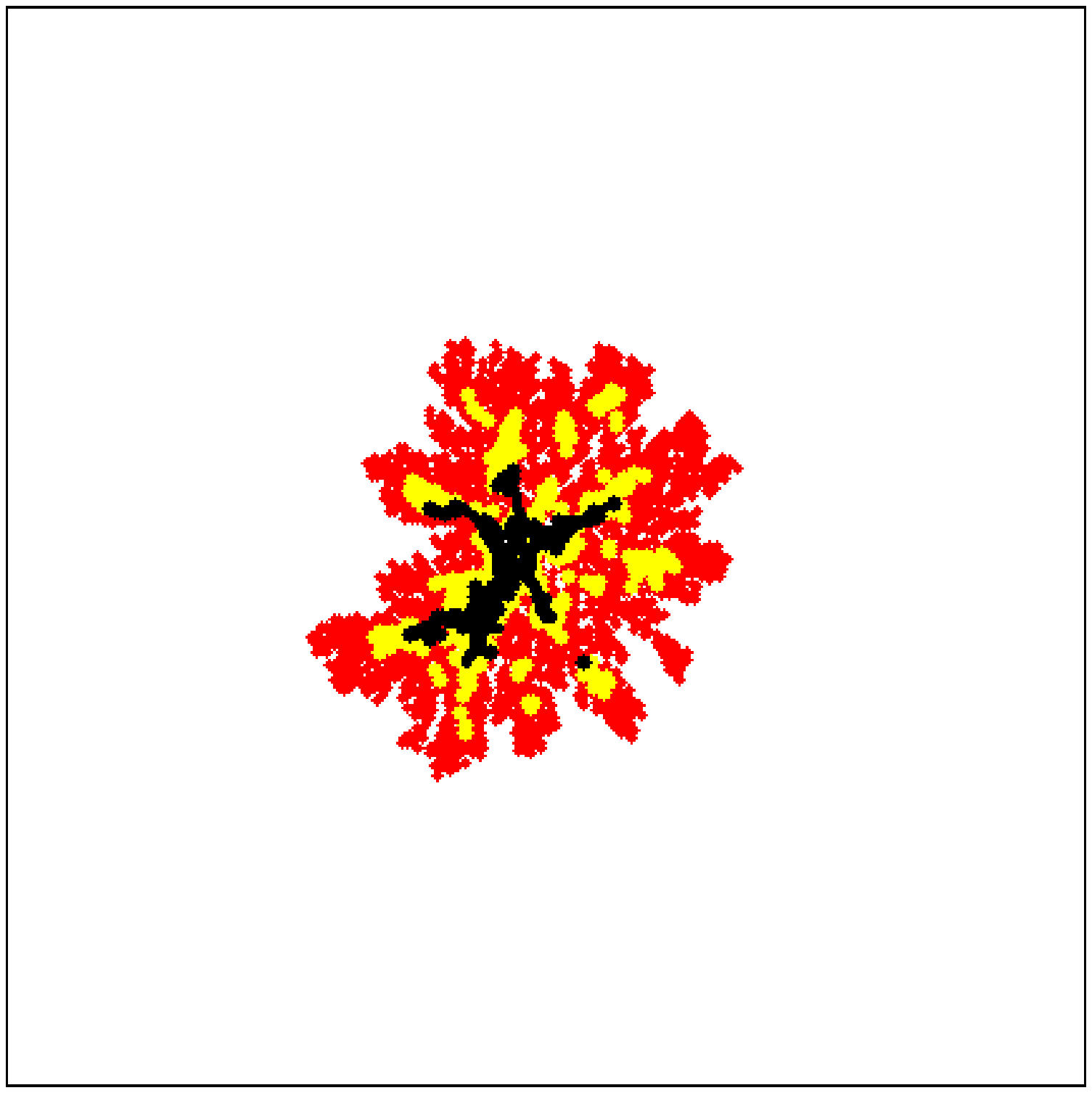} &
\includegraphics[height=3.5cm, keepaspectratio]{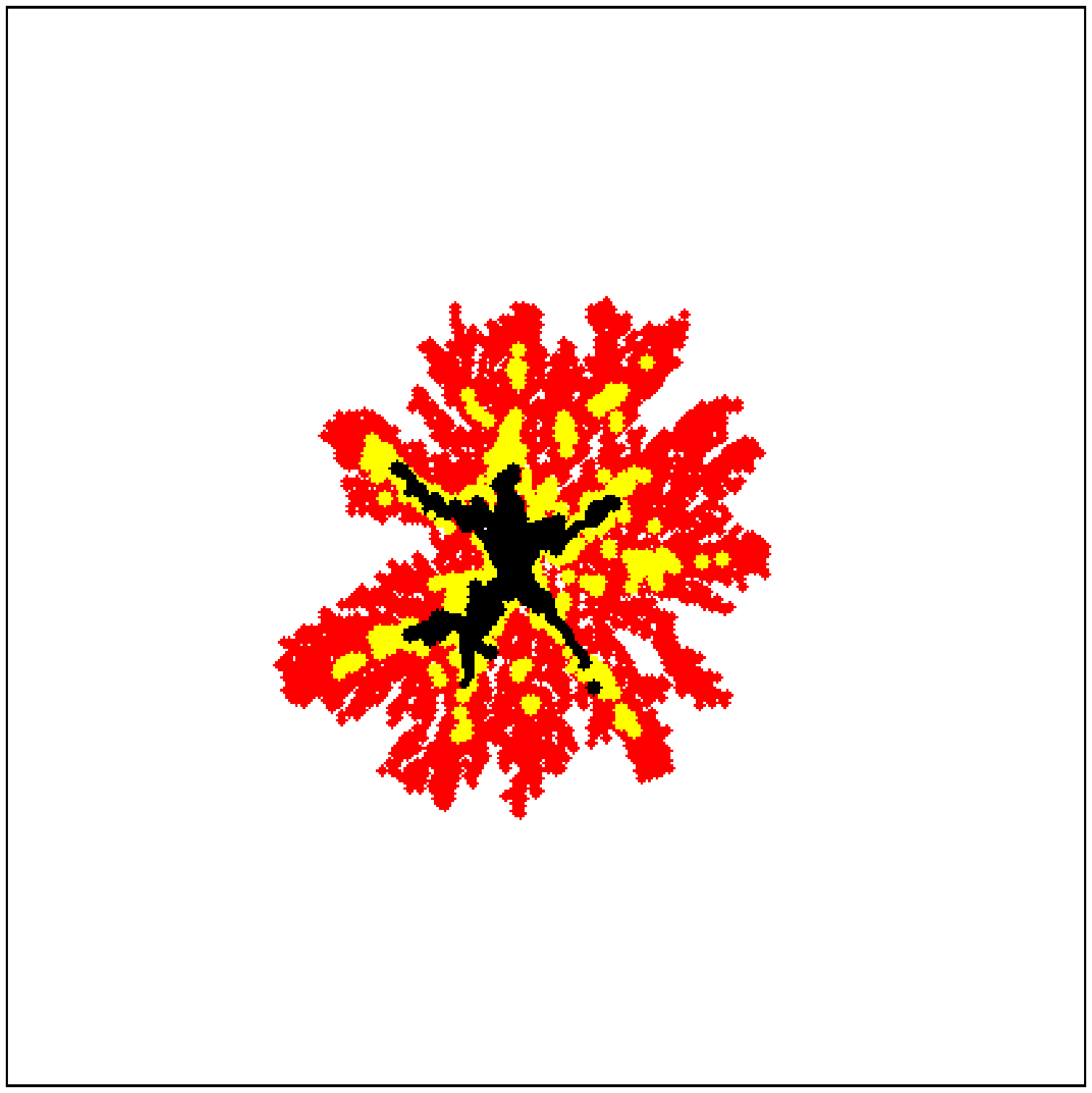} &
\includegraphics[height=3.5cm, keepaspectratio]{fig1c4.eps} \\
\mbox{(c1)} & \mbox{(c2)} & \mbox{(c3)} & \mbox{(c4)} \\\\
\end{array}$
\centering \caption{Jiao, Torquato}
\end{figure}

\clearpage
\newpage

\begin{figure}
$\begin{array}{c@{\hspace{0.5cm}}c@{\hspace{0.5cm}}c@{\hspace{0.5cm}}c}
\includegraphics[height=3.5cm, keepaspectratio]{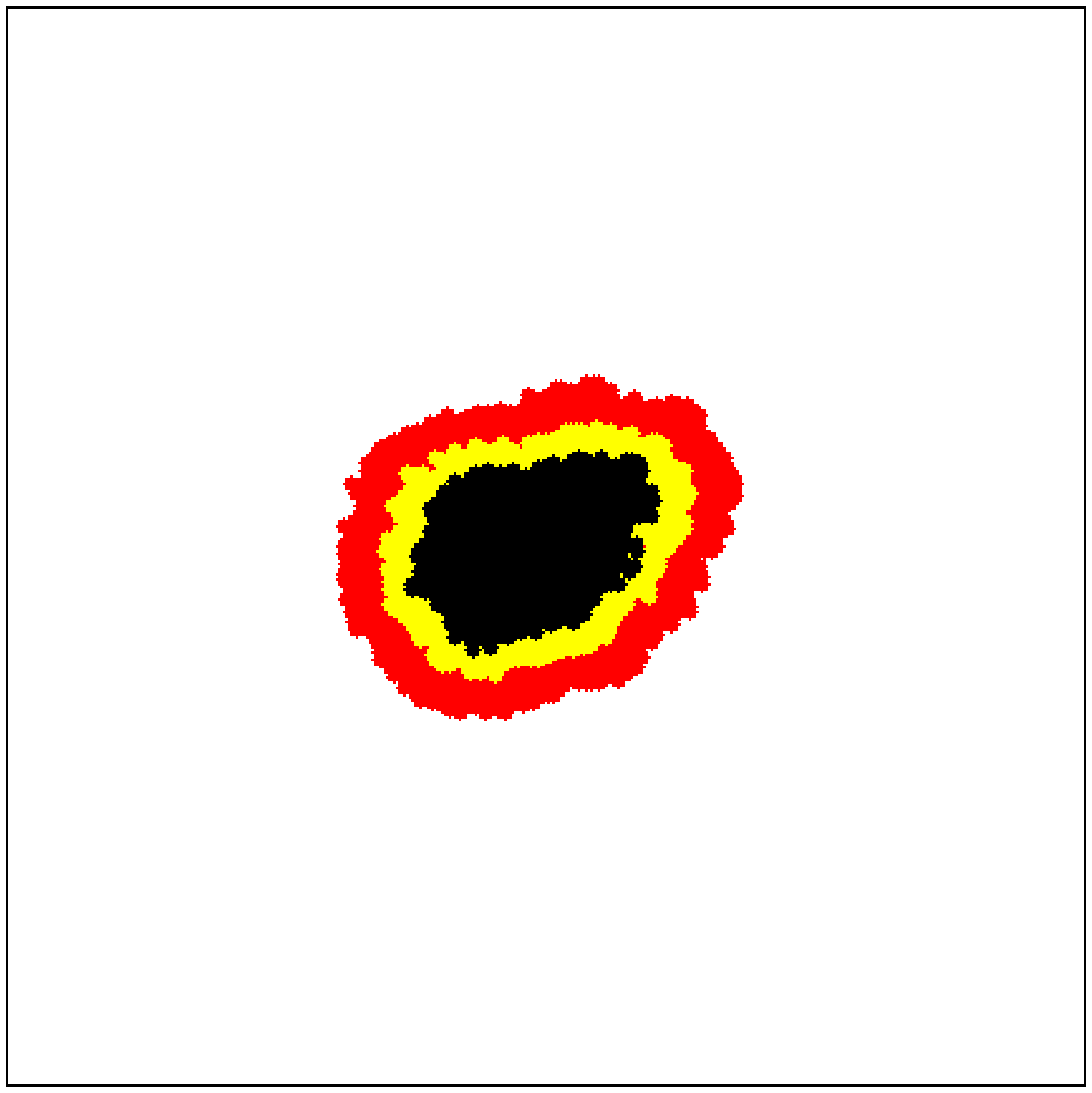} &
\includegraphics[height=3.5cm, keepaspectratio]{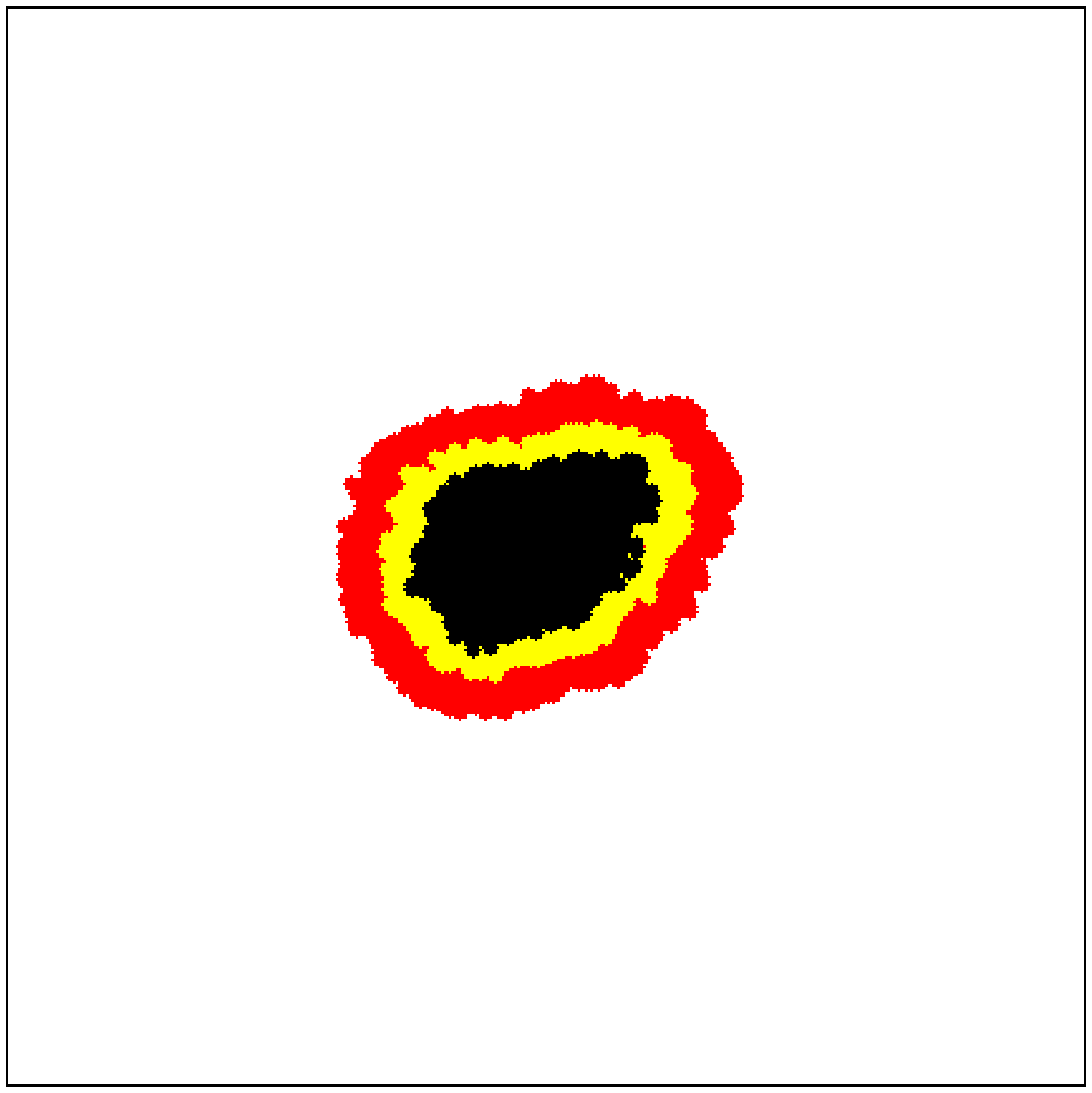} &
\includegraphics[height=3.5cm, keepaspectratio]{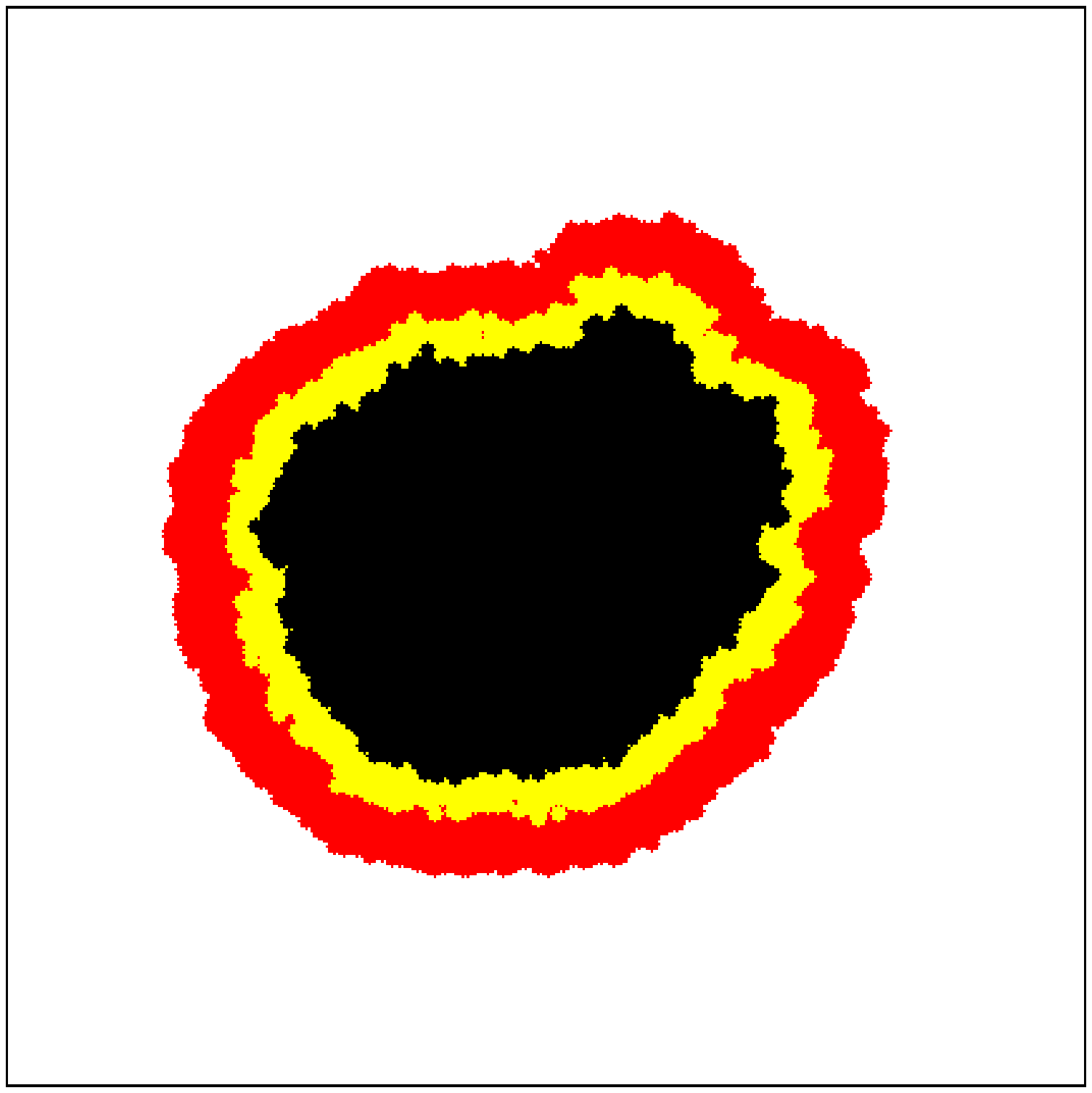} &
\includegraphics[height=3.5cm, keepaspectratio]{fig2a4.eps} \\
\mbox{(a1)} & \mbox{(a2)} & \mbox{(a3)} & \mbox{(a4)} \\\\
\includegraphics[height=3.5cm, keepaspectratio]{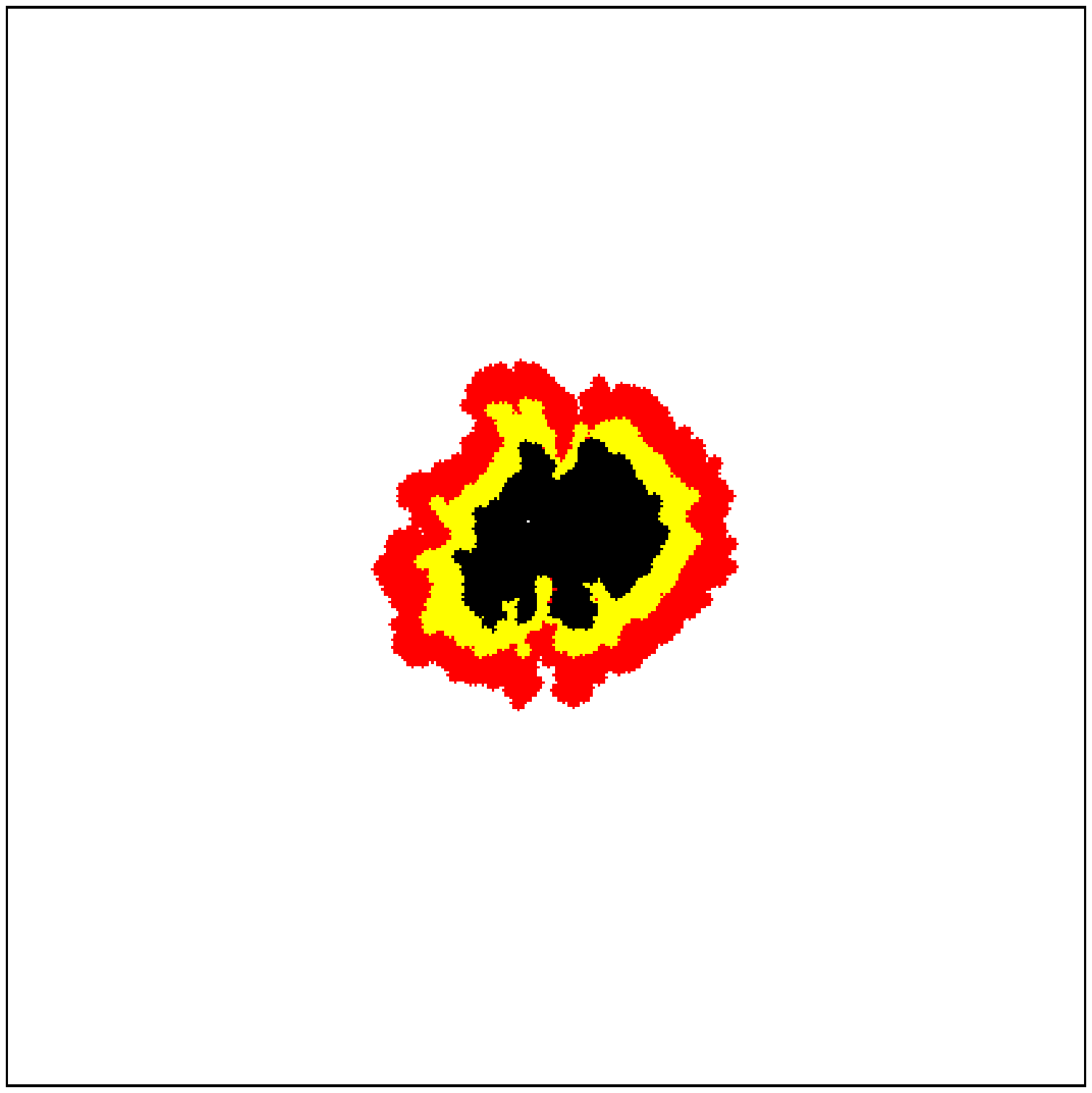} &
\includegraphics[height=3.5cm, keepaspectratio]{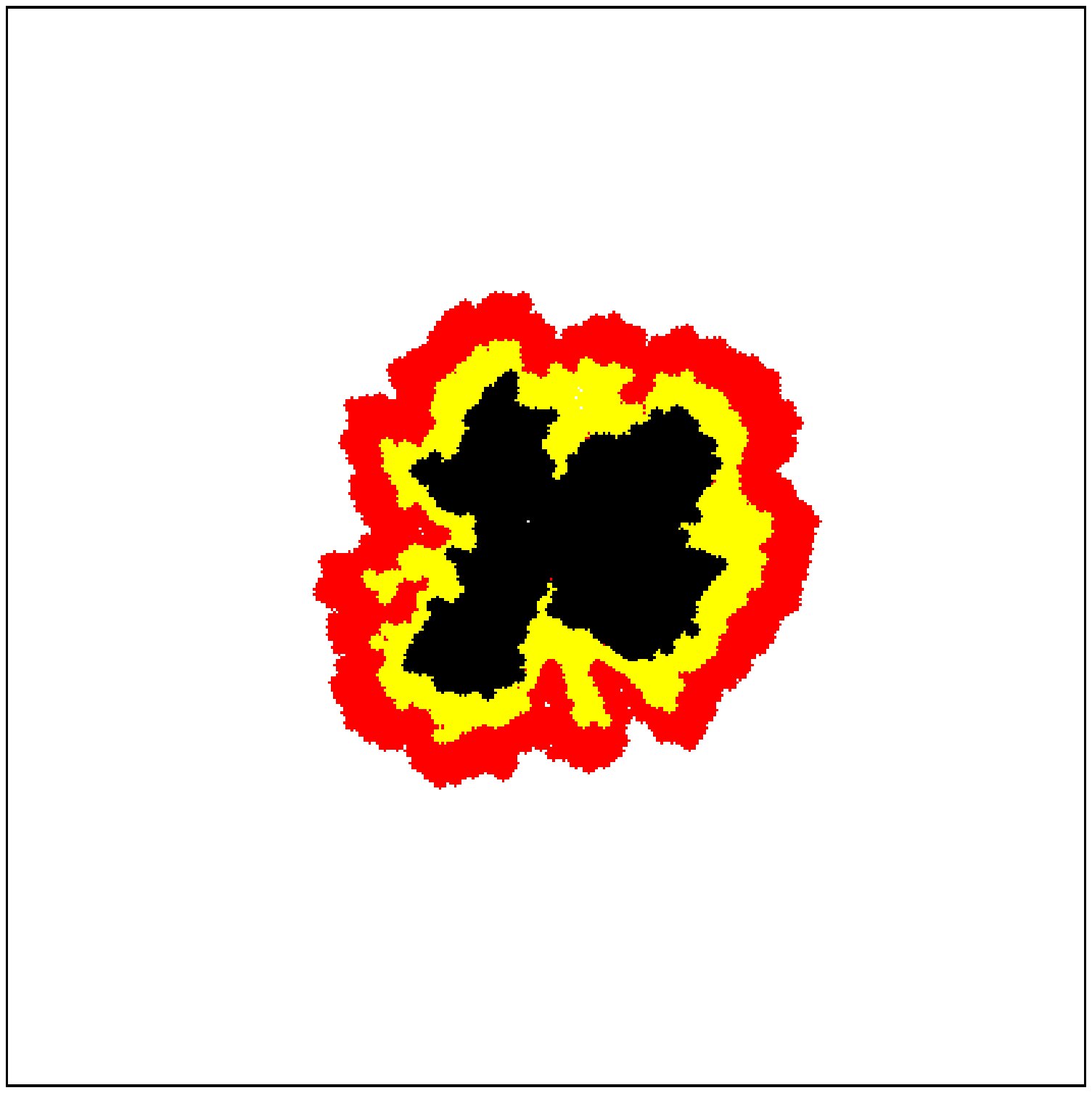} &
\includegraphics[height=3.5cm, keepaspectratio]{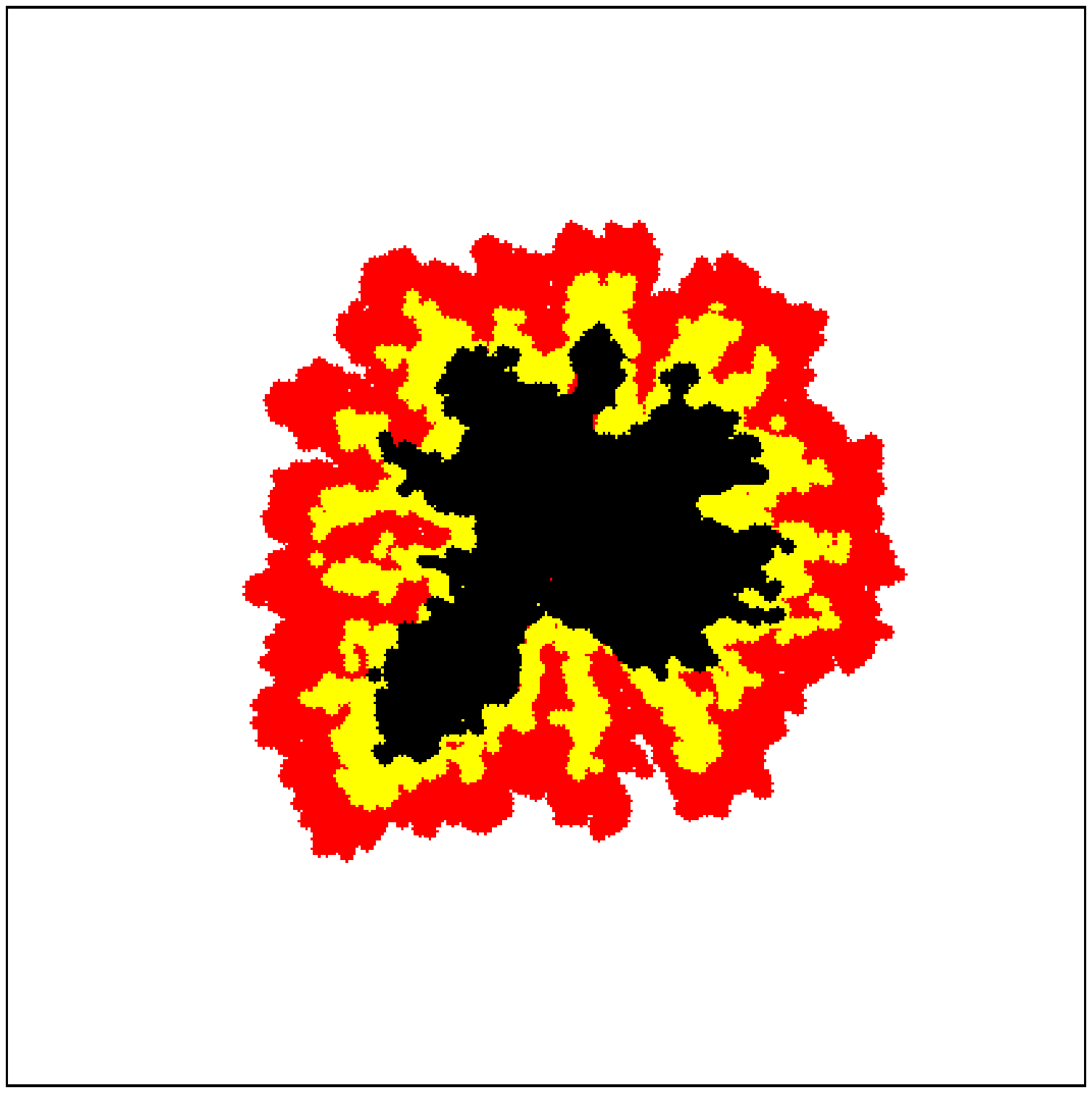} &
\includegraphics[height=3.5cm, keepaspectratio]{fig2b4.eps} \\
\mbox{(b1)} & \mbox{(b2)} & \mbox{(b3)} & \mbox{(b4)} \\\\
\includegraphics[height=3.5cm, keepaspectratio]{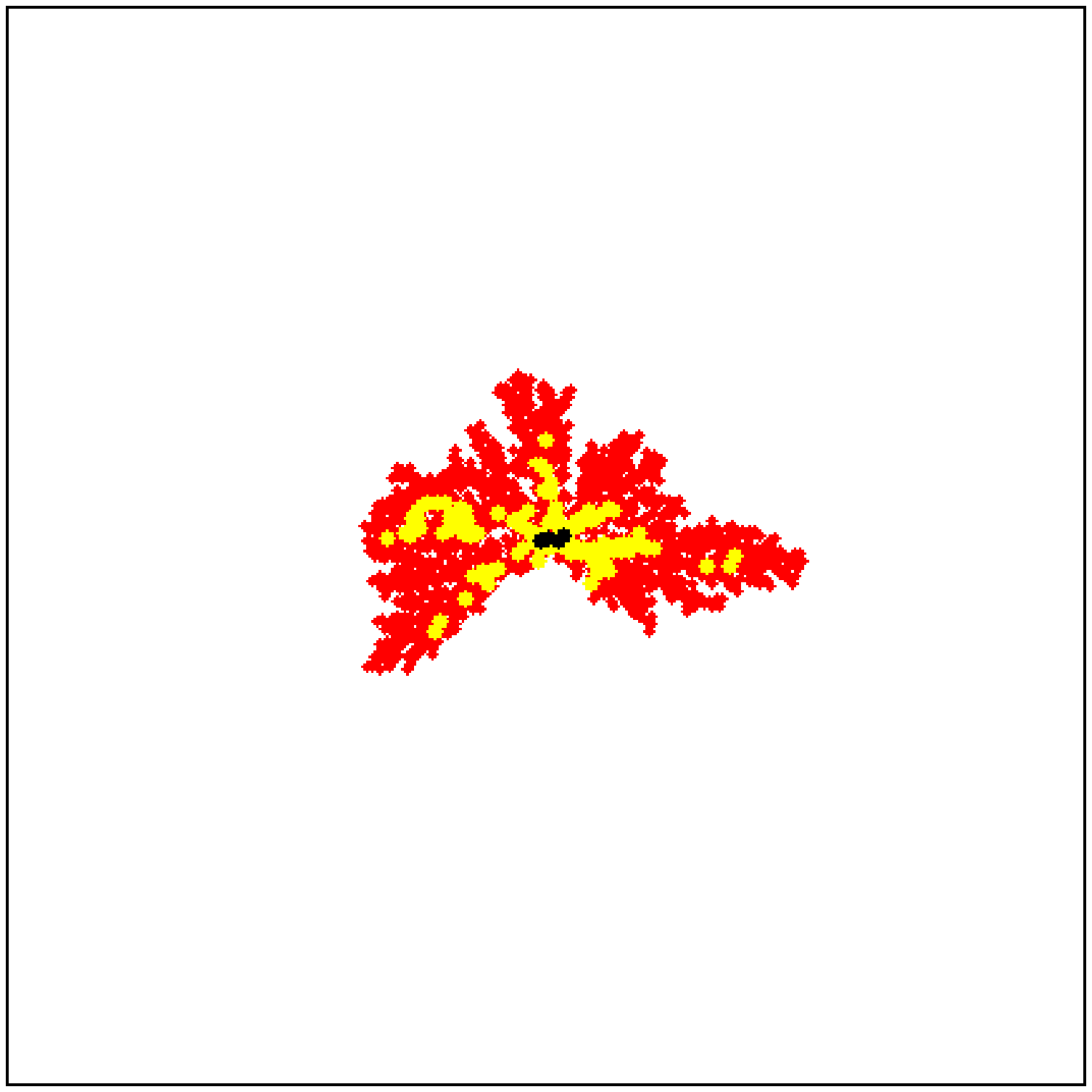} &
\includegraphics[height=3.5cm, keepaspectratio]{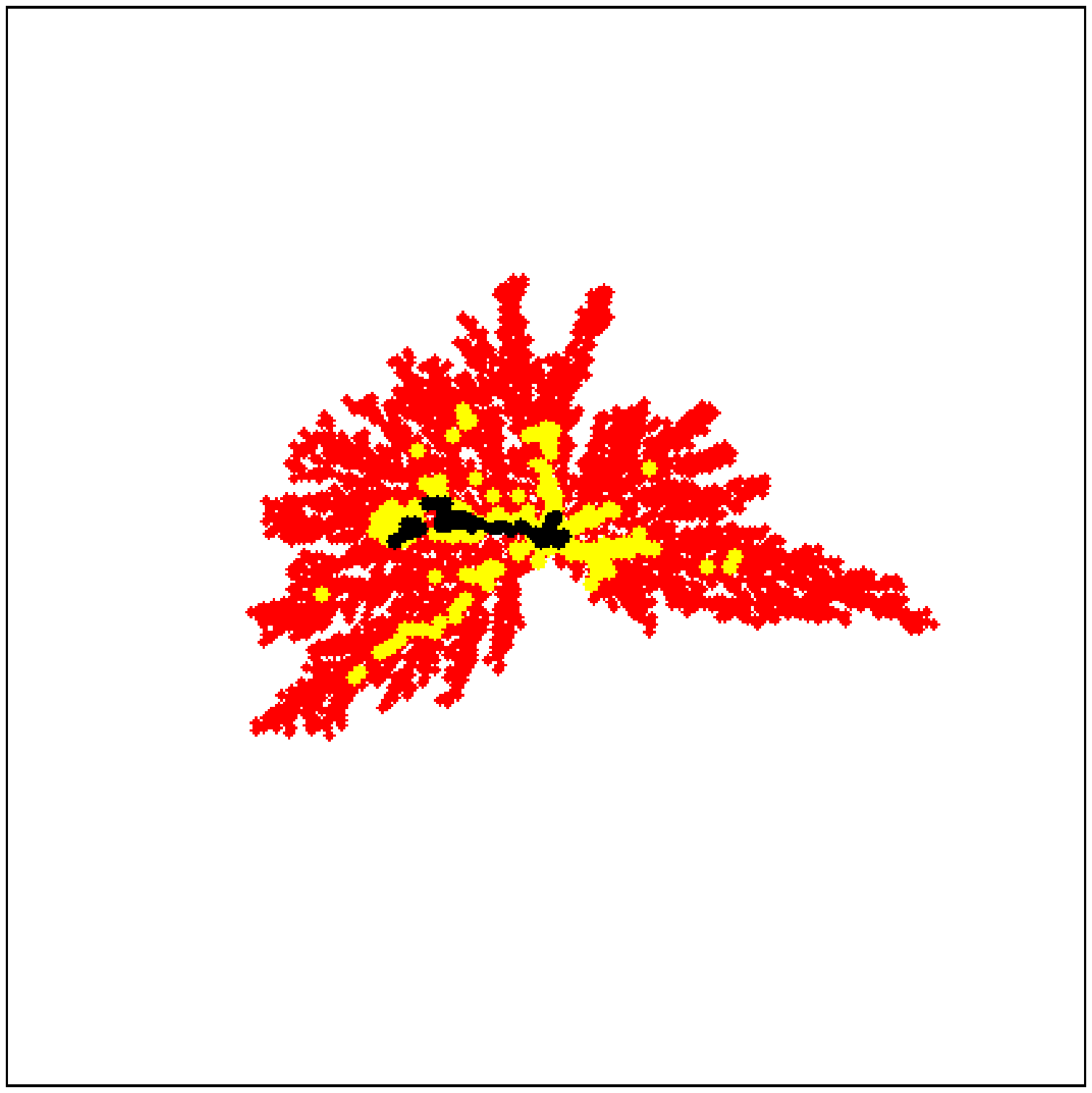} &
\includegraphics[height=3.5cm, keepaspectratio]{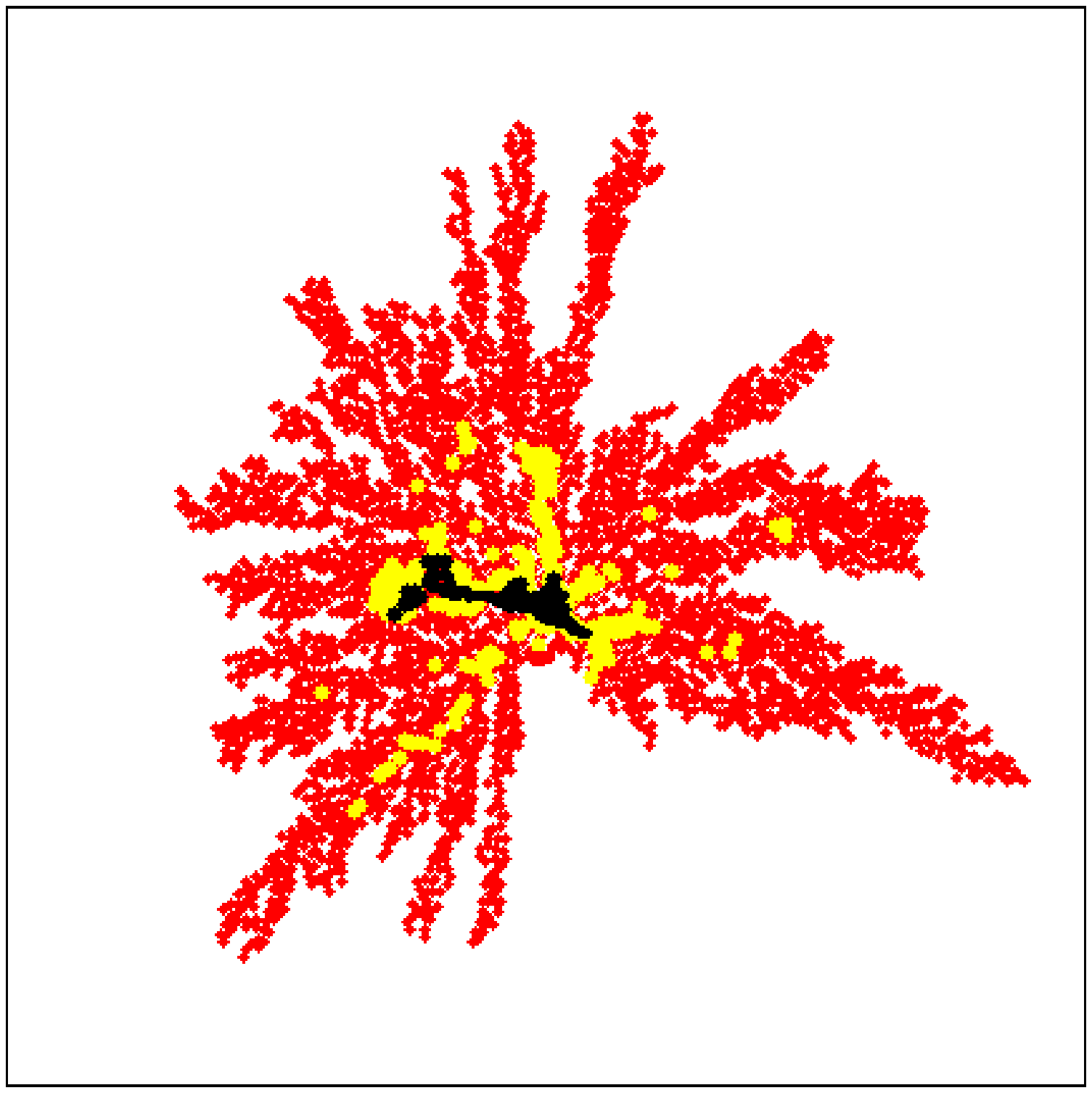} &
\includegraphics[height=3.5cm, keepaspectratio]{fig2c4.eps} \\
\mbox{(c1)} & \mbox{(c2)} & \mbox{(c3)} & \mbox{(c4)} \\\\
\end{array}$
\centering \caption{Jiao, Torquato}
\end{figure}

\clearpage
\newpage

\begin{figure}
$\begin{array}{c@{\hspace{0.5cm}}c@{\hspace{0.5cm}}c}
\includegraphics[height=4.5cm, keepaspectratio]{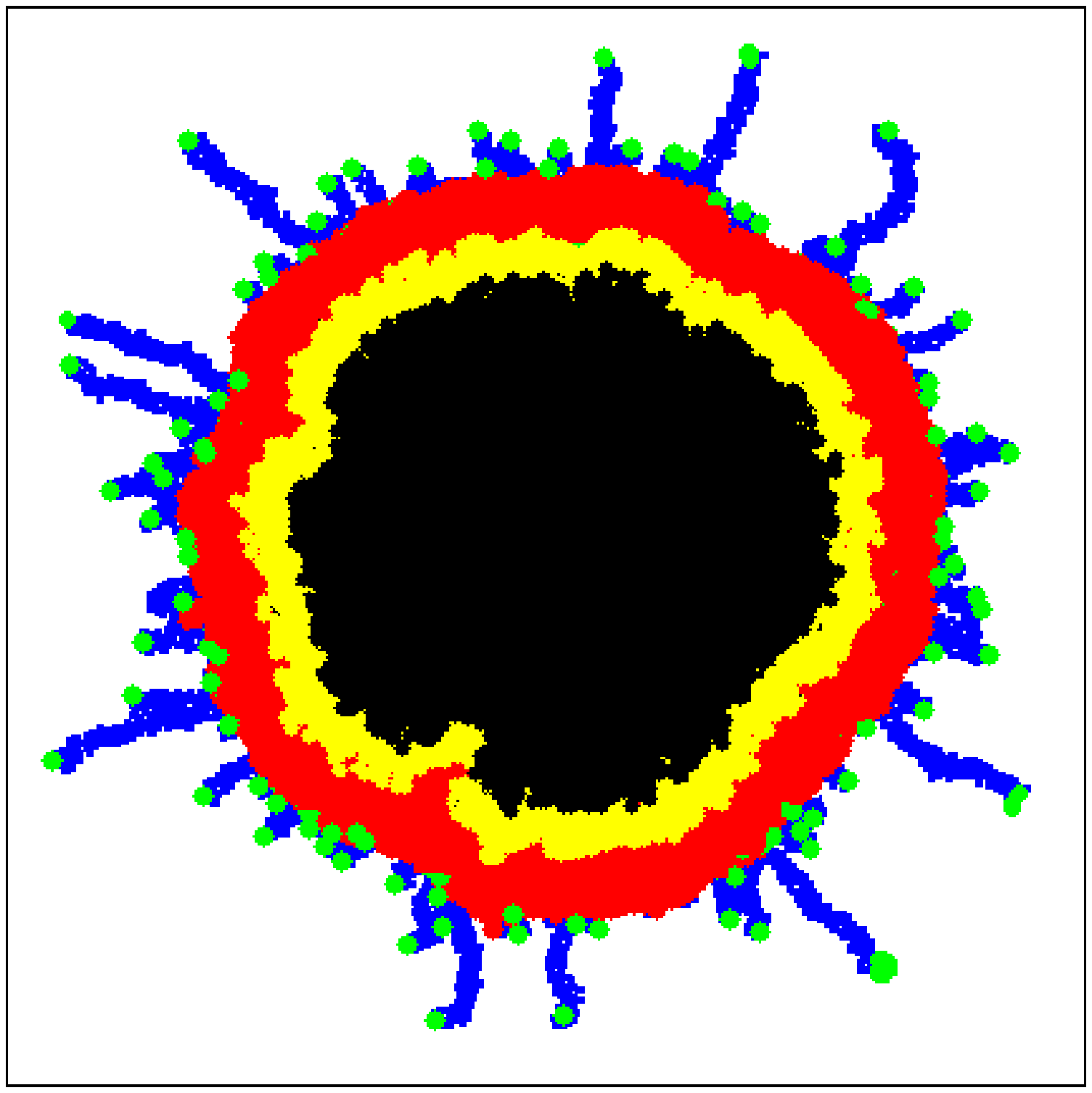} &
\includegraphics[height=4.5cm, keepaspectratio]{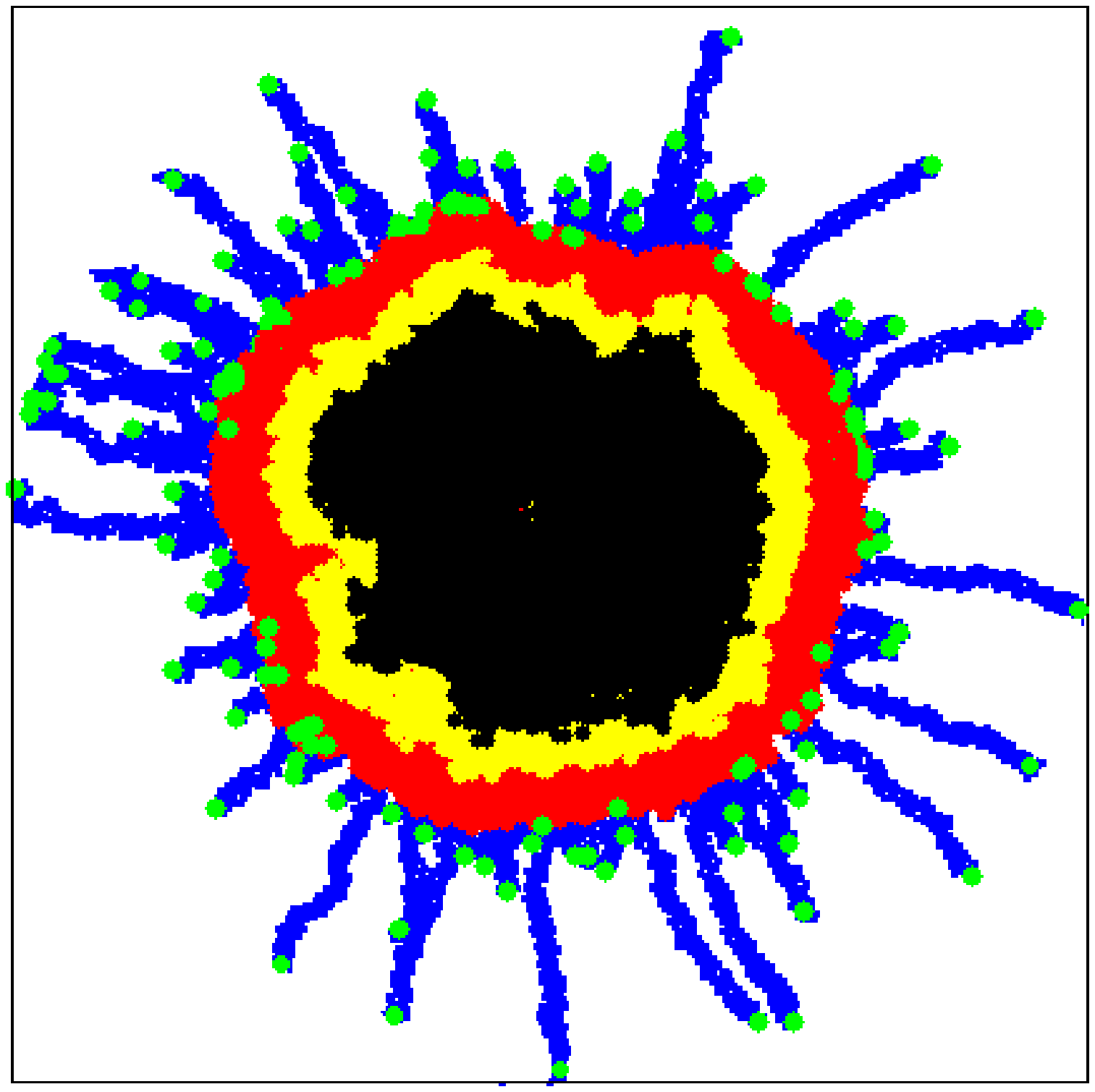} &
\includegraphics[height=4.5cm, keepaspectratio]{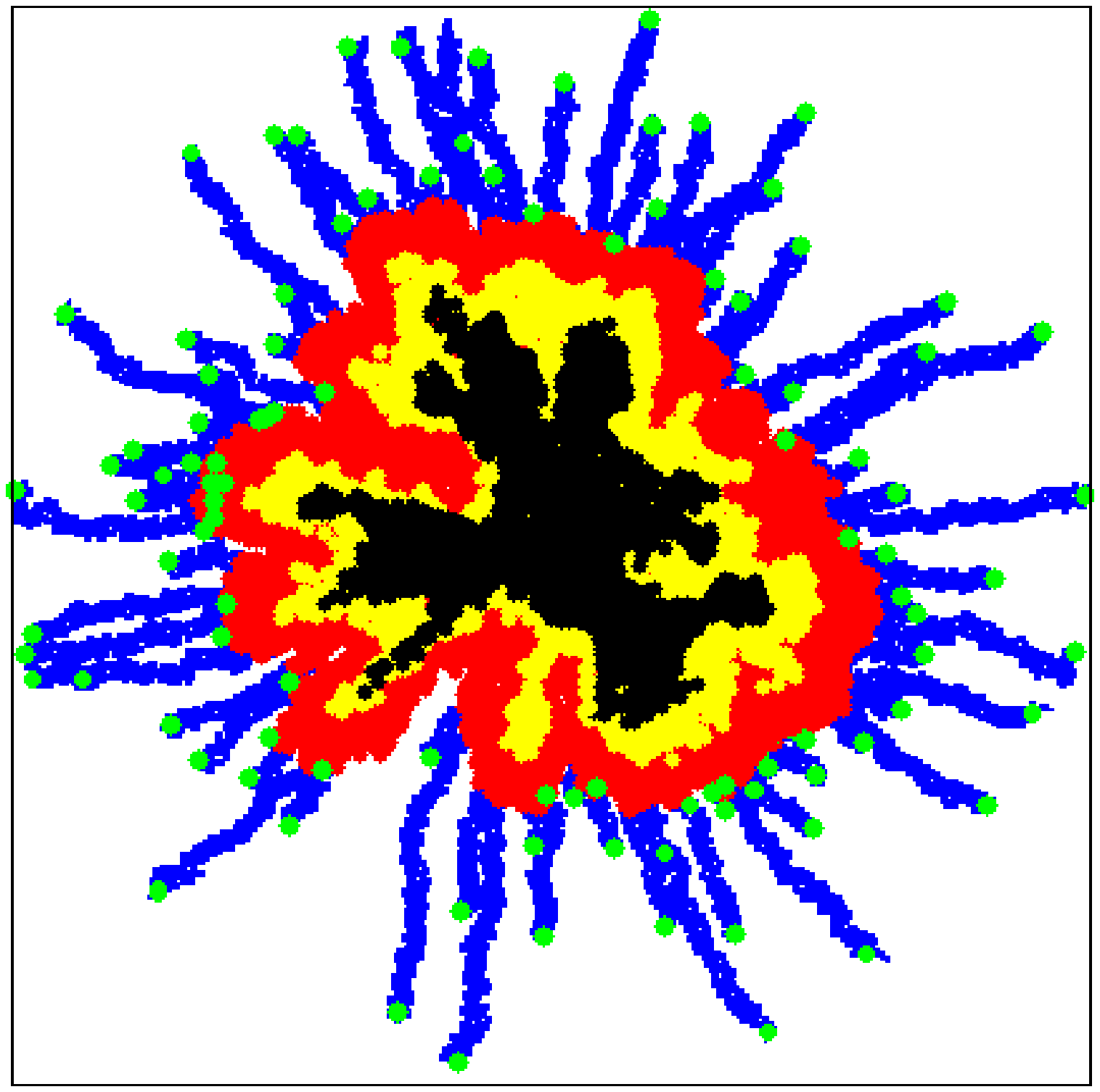} \\
\mbox{(a)} & \mbox{(b)} & \mbox{(c)} \\\\
\end{array}$
\centering \caption{Jiao, Torquato}
\end{figure}


\begin{figure}
$\begin{array}{c@{\hspace{0.5cm}}c@{\hspace{0.5cm}}c}
\includegraphics[height=4.5cm, keepaspectratio]{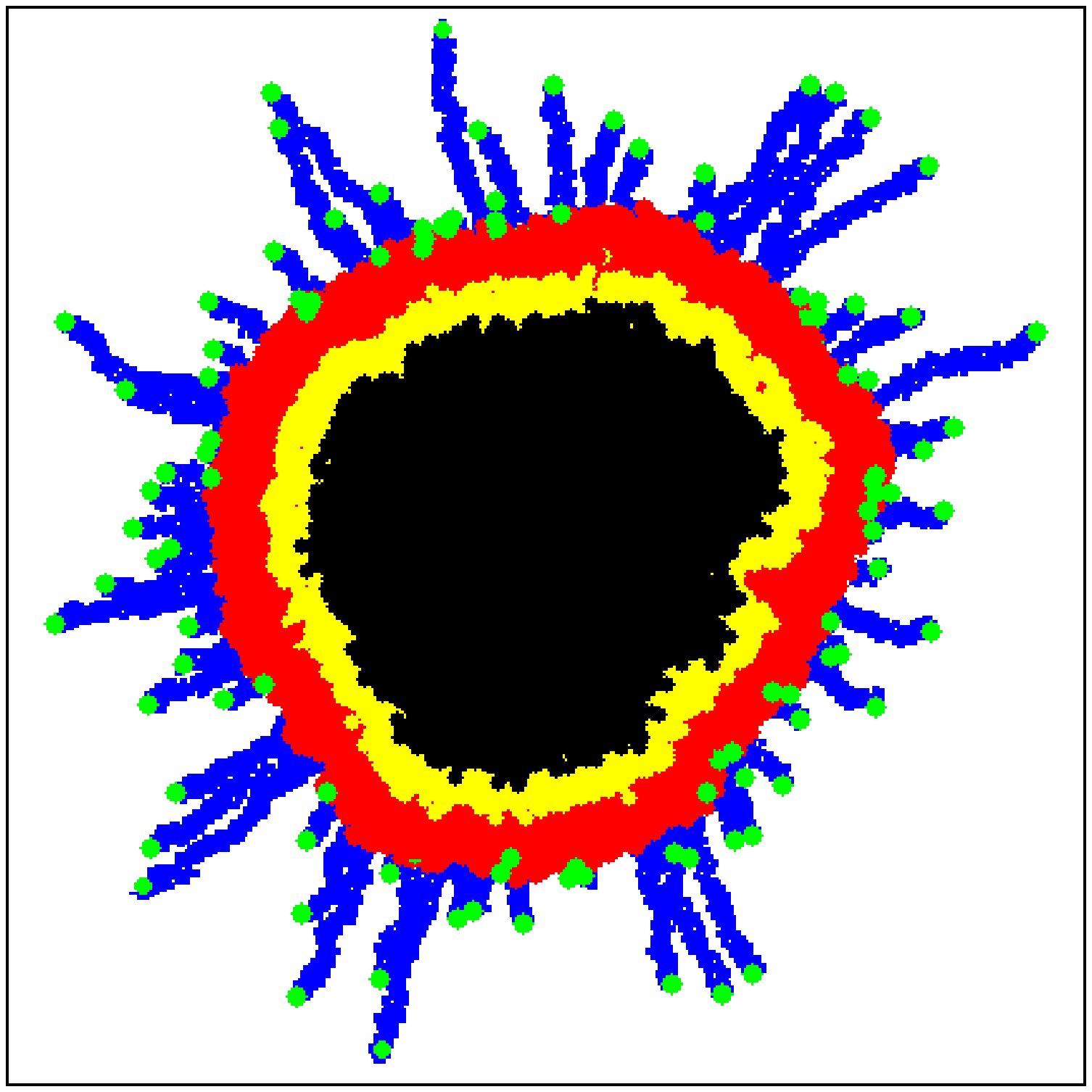} &
\includegraphics[height=4.5cm, keepaspectratio]{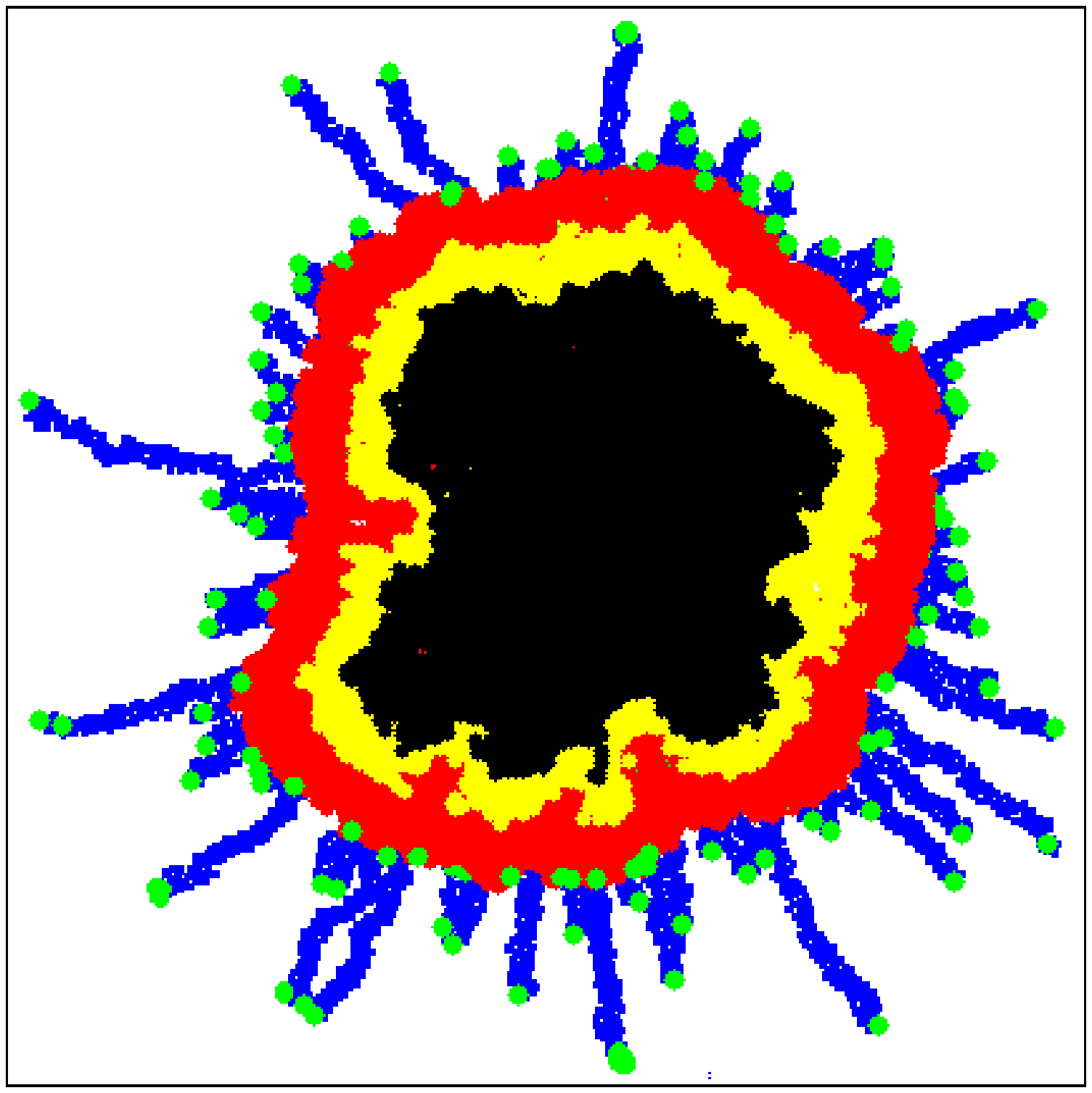} &
\includegraphics[height=4.5cm, keepaspectratio]{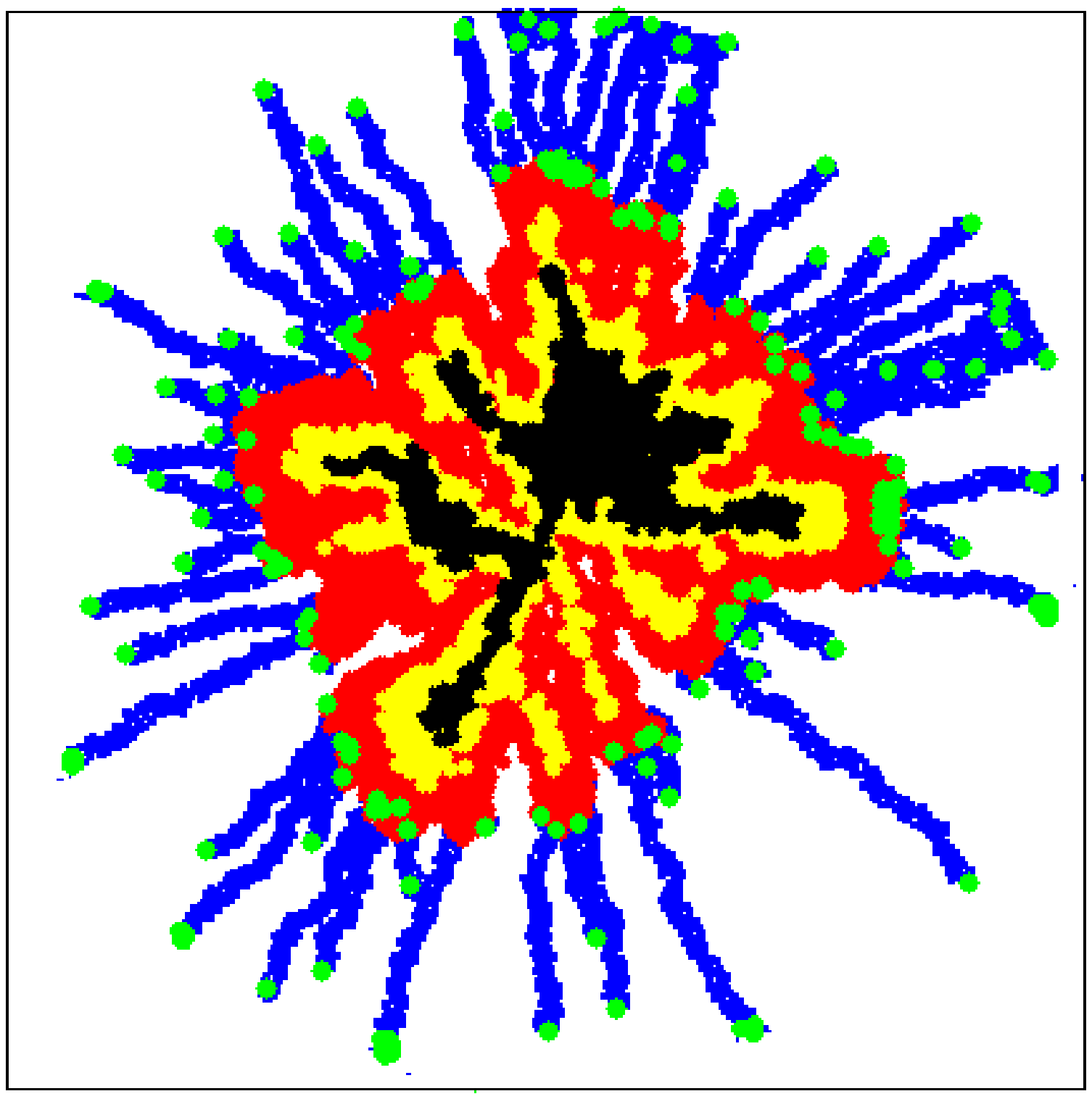} \\
\mbox{(a)} & \mbox{(b)} & \mbox{(c)} \\\\
\end{array}$
\centering \caption{Jiao, Torquato}
\end{figure}

\begin{figure}
$\begin{array}{c}
\includegraphics[height=3.5cm, keepaspectratio]{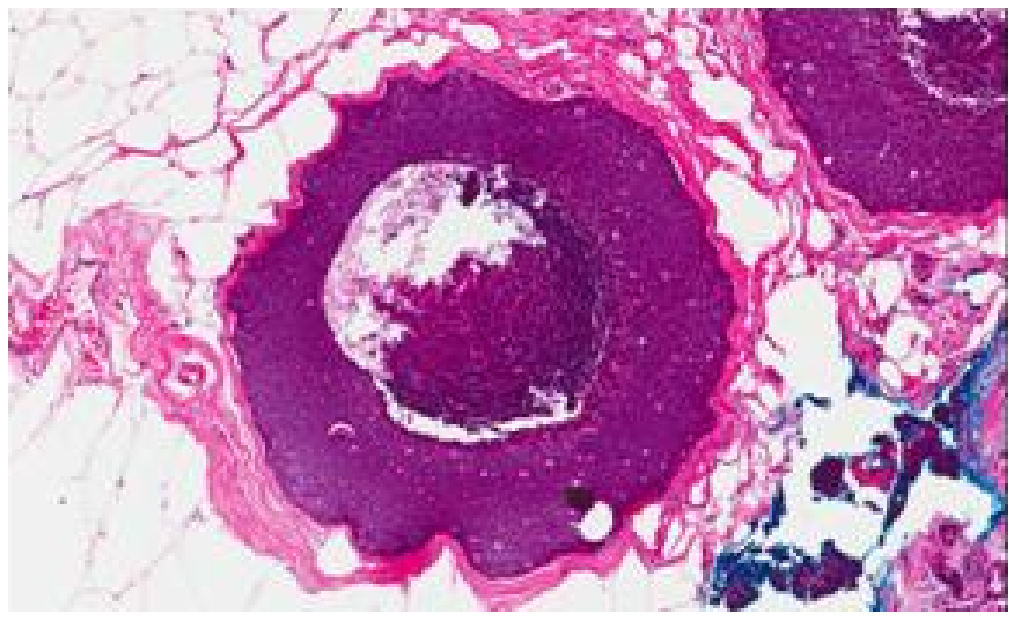} \\
\end{array}$
\centering \caption{Jiao, Torquato}
\end{figure}

\end{document}